\documentclass{aa}
\usepackage{graphicx}
\usepackage{txfonts}
\usepackage{natbib}
\begin{document}
   \title{Radial temperature profiles for a large sample of galaxy clusters observed with \emph{XMM-Newton}}


   \author{A. Leccardi
          \inst{1,2}
          \and
          S. Molendi\inst{2}
          }


   \institute{Universit\`a degli Studi di Milano, Dip. di Fisica, via Celoria 16,
              I-20133 Milano, Italy
         \and
             INAF-IASF Milano, via Bassini 15, I-20133 Milano, Italy
             }

   \date{Received February 8, 2008; accepted April 8, 2008}

 
  \abstract
   {}
   {We measure, as far out as possible, radial temperature profiles for a sample
of $\approx 50$ hot, intermediate redshift galaxy clusters, selected from the
\emph{XMM-Newton} archive, keeping systematic errors under control.
}
   {Our work is characterized by two major improvements. Firstly, we use the
background modeling, rather than the background subtraction, and the Cash statistic
rather than the $\chi^2$; this method requires a careful characterization of all
background components. Secondly, we assess in details systematic effects.
We perform two groups of test: prior to the analysis, we make use of extensive
simulations to quantify the impact of different spectral components on simulated
spectra; after the analysis, we investigate how the measured temperature
profile changes, when choosing different key parameters.
}
   {The mean temperature profile declines beyond 0.2~$R_{180}$; for the first
time we provide an assessment of the source and the magnitude of systematic uncertainties.
When comparing our profile with that obtained from hydrodynamic simulations, we
find the slopes beyond $\approx$~0.2~$R_{180}$ to be similar.
Our mean profile is similar but somewhat flatter with respect to that obtained
by previous observational works, possibly as a consequence of a different level of
characterization of systematic effects.
}
   {This work allows us not only to constrain with confidence cluster temperature
profiles in outer regions, but also, from a more general point of view, to explore
the limits of the current X-ray experiments (in particular \emph{XMM-Newton}) with
respect to the analysis of low surface brightness emission.
}


   \maketitle
%

\section{Introduction} \label{sec: intro}
Clusters of galaxies are the most massive gravitationally bound systems in the
universe.
They are permeated by the hot, X-ray emitting, intra-cluster medium (ICM),
which represents the dominant baryonic component.
The key ICM observable quantities are its density, temperature, and
metallicity.
Assuming hydrostatic equilibrium, the gas temperature and density profiles
allow us to derive the total cluster mass and thus to use galaxy clusters as
cosmological probes (e.g. \citealp{henry91,ettori02,fabian03c,voit05}).
Temperature and density profiles can also be combined to determine the ICM
entropy distribution, that provides valuable information on the cluster
thermodynamic history and has proven to be a powerful tool to investigate
non-gravitational processes (e.g. \citealp{ponman03,mccarthy04,voit05,pratt06}).

Cluster outer regions are rich of information and interesting to study,
because clusters are still forming there by accretion
(e.g. \citealp{tozzi00,borgani04}); moreover, far from the core it is easier to
compare simulations with observations, because feedback effects are less
important (e.g. \citealp{borgani04,mcnamara05,ronca06}).
Cluster surface brightness rapidly declines with radius, while background (of
instrumental, solar, local, and cosmic origin) is roughly constant over the
detector.
For this reason, spectra accumulated in the outer regions are characterized by
poor statistics and high background, especially at high energies, where
the instrumental background dominates other components.
These conditions make temperature measurement at large distances from the
center a technically challenging task, requiring an adequate treatment of both
statistical and systematic issues \citep{leccardi07}.

Given the technical difficulties, early measurements of cluster temperature
profiles have been controversial.
At the end of the \emph{ASCA} and \emph{BeppoSAX} era, the shape of the
profiles at large radii was still the subject of debate \citep{mark98b,irwin99,
white00,irwin00,fino01,grandi02}.
Recent observations with current experiments (i.e. \emph{XMM-Newton} and
\emph{Chandra}) have clearly shown that cluster temperature profiles decline
beyond the 15-20\% of R$_{180}$ \citep{piffa05,vikh05,pratt07,snowden08}.
However, most of these measurements might be unreliable at very large radii
($\gtrsim$~50\% of R$_{180}$) because they are affected by a number
of systematics related to the analysis technique and to the background
treatment \citep{leccardi07}.

The aim of this work is to measure the mean temperature profile of galaxy
clusters as far out as possible, while keeping systematic errors under control.
We select from the \emph{XMM-Newton} archive all hot ($\mathrm{k}T > 3.5$~keV),
intermediate redshift ($0.1 \lesssim z \lesssim 0.3$) clusters, that are not
strongly interacting, and measure their radial temperature profiles.
The spectral analysis follows a new approach: we use the background
modeling, rather than the background subtraction, and the Cash statistic
rather than the $\chi^2$.
This method requires a careful characterization (reported in the Appendices)
of all background components, which unfortunately has not been possible for
EPIC-pn; for this reason, in our analysis we use only EPIC-MOS data.

Background parameters are estimated in a peripheral region, where the cluster
emission is almost negligible, and rescaled in the regions of interest.
The spectral fitting is performed in the 0.7-10.0~keV and in the 2.0-10.0~keV
energy bands, that are characterized by different statistics and level of
systematics, to check the consistency of our results.
A second important point is a particular attention to systematic effects.
We perform two groups of test: prior to the analysis, we make use of extensive
simulations to quantify the impact of different components (e.g. the cosmic
variance or the soft proton contribution) on simulated spectra; after the
analysis, we investigate how the measured temperature profile changes, when
choosing different key parameters (e.g. the truncation radius or the energy
band).
At the end of our tests, we provide an assessment of the source and the
magnitude of systematic uncertainties associated to the mean profile.

We compare our profiles with those obtained from hydrodynamic simulations
\citep{borgani04} and from previous observational works
\citep{grandi02,vikh05,pratt07}.
Our work does not only provide a confirmation of previous results.
For the first time we believe we know where the systematics come from and
how large they are.
Indeed, this work allows us not only to constrain with confidence
cluster temperature profiles in the outer regions, but also, from a more
general point of view, to explore the limits of the current X-ray experiments
(in particular \emph{XMM-Newton}).
It is crucial that we learn how best to exploit \emph{XMM-Newton} data,
because for the next 5-10 years there will be no experiments with comparable
or improved capabilities, as far as low surface brightness emission is concerned.
Our work will also allow us to look forward to ambitious new measurements:
an example is the attempt to measure the putative shock in Abell~754, for
which we have obtained a $\approx 200$~ks observation with \emph{XMM-Newton}
in AO7.

The outline of the paper is the following.
In Sect.~\ref{sec: sample} we describe sample properties and selection
criteria and in Sect.~\ref{sec: analysis} we describe in detail our data
analysis technique.
In Sect.~\ref{sec: profiles} we present the radial temperature profiles for
all clusters in our sample and compute the average profile.
In Sect.~\ref{sec: syst} we describe our analysis of systematic effects.
In Sect.~\ref{sec: mean prof} we characterize the profile decline, investigate
its dependency from physical properties (e.g. the redshift), and compare
it with hydrodynamic simulations and previous observational works.
Our main results are summarized in Sect.~\ref{sec: concl}.
In the Appendices we report the analysis of closed and blank field
observations, which allows us to characterize most background components.

Quoted confidence intervals are 68\% for one interesting parameter
(i.e. $\Delta$C~=~1), unless otherwise stated.
All results are given assuming a $\Lambda$CDM cosmology with
$\Omega_\mathrm{m} = 0.3$, $\Omega_\Lambda = 0.7$, and
$H_0$~=~70~km~s$^{-1}$~Mpc$^{-1}$.


\section{The sample}  \label{sec: sample}
We select from the \emph{XMM-Newton} archive a sample of hot
($\mathrm{k}T > 3.3$~keV), intermediate redshift ($0.1 \lesssim z \lesssim 0.3$),
and high galactic latitude ($|b| > 20^\circ$) clusters of galaxies.
Upper and lower limits to the redshift range are determined, respectively,
by the cosmological dimming effect and the size of the EPIC field of view
($\approx 15^\prime$ radius).
Indeed, our data analysis technique requires that the intensity of
background components be estimated in a peripheral region, where the
cluster emission is almost negligible (see Sect.~\ref{sec: ext ring}).
We retrieve from the public archive all observations of clusters satisfying
the above selection criteria, performed before March 2005 (when the CCD6
of EPIC-MOS1 was switched off\footnote{http://xmm.vilspa.esa.es/external/xmm\_news/items/MOS1-CCD6/\\
index.shtml}) and available at the end of May 2007.
\begin{table}
  \caption{Observations excluded from the sample due to high soft proton
  contamination.}
  \label{tab: excl SP}
  \centering
  \begin{tabular}{ll}
    \hline \hline
    Name & Obs ID \\
    \hline
    RXCJ0303.8-7752	& 0042340401 \\
    RXCJ0516.7-5430	& 0042340701 \\
    RXCJ0528.9-3927	& 0042340801 \\
    RXCJ2011.3-5725	& 0042341101 \\
    Abell 2537		& 0042341201 \\
    RXCJ0437.1+0043	& 0042341601 \\
    Abell 1302		& 0083150401 \\
    Abell 2261		& 0093030301 \\
    Abell 2261		& 0093030801 \\
    Abell 2261		& 0093030901 \\
    Abell 2261		& 0093031001 \\
    Abell 2261		& 0093031101 \\
    Abell 2261		& 0093031401 \\
    Abell 2261		& 0093031501 \\
    Abell 2261		& 0093031601 \\
    Abell 2261		& 0093031801 \\
    Abell 2219		& 0112231801 \\
    Abell 2219		& 0112231901 \\
    RXCJ0006.0-3443	& 0201900201 \\
    RXCJ0145.0-5300	& 0201900501 \\
    RXCJ0616.8-4748	& 0201901101 \\
    RXCJ0437.1+0043	& 0205330201 \\
    Abell 2537		& 0205330501 \\
    \hline
  \end{tabular}
\end{table}
Unfortunately, 23 of these 86 observations are highly
affected by soft proton flares (see Table~\ref{tab: excl SP}).
We exclude them from the sample, because their good (i.e. after
flare cleaning, see Sect.~\ref{sec: prelim}) exposure time is not 
sufficient (less than 16~ks when summing MOS1 and MOS2) to measure
reliable temperature profiles out to external regions.
\begin{table}
  \caption{Observations of clusters that show evidence of recent
  and strong interactions.}
  \label{tab: excl int}
  \centering
  \begin{tabular}{ll}
    \hline \hline
    Name & Obs ID \\
    \hline
    Abell 2744		& 0042340101 \\
    Abell 665		& 0109890401 \\
    Abell 665		& 0109890501 \\
    Abell 1914		& 0112230201 \\
    Abell 2163		& 0112230601 \\
    Abell 2163		& 0112231501 \\
    RXCJ0658.5-5556	& 0112980201 \\
    Abell 1758		& 0142860201 \\
    Abell 1882		& 0145480101 \\
    Abell 901		& 0148170101 \\
    Abell 520		& 0201510101 \\
    Abell 2384		& 0201902701 \\
    Abell 115		& 0203220101 \\
    ZwCl2341.1+0000	& 0211280101 \\
    \hline
  \end{tabular}
\end{table}
Furthermore, we exclude 14 observations of clusters that show evidence
of recent and strong interactions (see Table~\ref{tab: excl int}).
For such clusters, a radial analysis is not appropriate, because
the gas distribution is far from being azimuthally symmetric.
Finally, we find that the target of observation 0201901901, which
is classified as a cluster, is likely a point-like source; therefore,
we exclude this observation too from our sample.

\begin{table*}
  \caption{Physical properties and observation details for the 48
  clusters of our sample.}
  \label{tab: sample}
  \centering
  \begin{tabular}{l l c r r r r r l}
    \hline \hline
    Name & Obs ID & $z\mathrm{^a}$ & $\mathrm{k}T_\mathrm{M}\mathrm{^b}$ &
    $R_{180}\mathrm{^c}$ & \multicolumn{2}{c}{Exp. time$\mathrm{^d}$} &
    $R_\mathrm{SB}\mathrm{^e}$ & Filter \\
    \hline
    RXCJ0043.4-2037	& 0042340201 & 0.2924 &  6.8 & 1.78 & 11.9 & 11.3 & 1.25 & THIN1 \\
    RXCJ0232.2-4420	& 0042340301 & 0.2836 &  7.2 & 1.85 & 12.1 & 11.7 & 1.08 & THIN1 \\
    RXCJ0307.0-2840	& 0042340501 & 0.2534 &  6.8 & 1.82 & 11.4 & 12.6 & 1.08 & THIN1 \\
    RXCJ1131.9-1955	& 0042341001 & 0.3072 &  8.1 & 1.93 & 12.4 & 12.3 & 1.08 & THIN1 \\
    RXCJ2337.6+0016	& 0042341301 & 0.2730 &  7.2 & 1.86 & 13.4 & 13.1 & 1.19 & THIN1 \\
    RXCJ0532.9-3701	& 0042341801 & 0.2747 &  7.5 & 1.90 & 10.9 & 10.5 & 1.09 & THIN1 \\
    Abell 68    	& 0084230201 & 0.2550 &  7.2 & 1.88 & 26.3 & 25.9 & 1.37 & MEDIUM \\
    Abell 209   	& 0084230301 & 0.2060 &  6.6 & 1.85 & 17.9 & 17.8 & 1.19 & MEDIUM \\
    Abell 267   	& 0084230401$^*$ & 0.2310 &  4.5 & 1.49 & 17.0 & 16.5 & 1.79 & MEDIUM \\
    Abell 383   	& 0084230501 & 0.1871 &  4.4 & 1.52 & 29.3 & 29.8 & 1.33 & MEDIUM \\
    Abell 773   	& 0084230601 & 0.2170 &  7.5 & 1.96 & 13.6 & 15.5 & 1.16 & MEDIUM \\
    Abell 963   	& 0084230701 & 0.2060 &  6.5 & 1.83 & 24.4 & 26.0 & 1.19 & MEDIUM \\
    Abell 1763  	& 0084230901 & 0.2230 &  7.2 & 1.92 & 13.0 & 13.2 & 1.08 & MEDIUM \\
    Abell 1689  	& 0093030101 & 0.1832 &  9.2 & 2.21 & 36.8 & 36.8 & 1.14 & THIN1 \\
    RX J2129.6+0005	& 0093030201 & 0.2350 &  5.5 & 1.66 & 36.0 & 37.5 & 1.21 & MEDIUM \\
    ZW 3146		& 0108670101 & 0.2910 &  7.0 & 1.81 & 52.9 & 52.9 & 1.07 & THIN1 \\
    E1455+2232  	& 0108670201 & 0.2578 &  5.0 & 1.56 & 35.3 & 35.8 & 1.11 & MEDIUM \\
    Abell 2390  	& 0111270101 & 0.2280 & 11.2 & 2.37 &  9.9 & 10.3 & 1.11 & THIN1 \\
    Abell 2204  	& 0112230301 & 0.1522 &  8.5 & 2.16 & 18.2 & 19.5 & 1.06 & MEDIUM \\
    Abell 1413  	& 0112230501 & 0.1427 &  6.7 & 1.92 & 25.4 & 25.4 & 1.10 & THIN1 \\
    Abell 2218  	& 0112980101 & 0.1756 &  6.5 & 1.86 & 18.2 & 18.2 & 1.17 & THIN1 \\
    Abell 2218  	& 0112980401 & 0.1756 &  7.0 & 1.93 & 13.7 & 14.0 & 1.42 & THIN1 \\
    Abell 2218  	& 0112980501 & 0.1756 &  6.1 & 1.80 & 11.3 & 11.0 & 1.07 & THIN1 \\
    Abell 1835  	& 0147330201 & 0.2532 &  8.6 & 2.05 & 30.1 & 29.2 & 1.16 & THIN1 \\
    Abell 1068  	& 0147630101 & 0.1375 &  4.5 & 1.58 & 20.5 & 20.8 & 1.09 & MEDIUM \\
    Abell 2667  	& 0148990101 & 0.2300 &  7.7 & 1.96 & 21.9 & 21.6 & 1.48 & MEDIUM \\
    Abell 3827  	& 0149670101 & 0.0984 &  7.1 & 2.02 & 22.3 & 22.4 & 1.16 & MEDIUM \\
    Abell 3911  	& 0149670301 & 0.0965 &  5.4 & 1.77 & 25.8 & 26.1 & 1.43 & THIN1 \\
    Abell 2034  	& 0149880101 & 0.1130 &  7.0 & 1.99 & 10.2 & 10.5 & 1.16 & THIN1 \\
    RXCJ0003.8+0203	& 0201900101 & 0.0924 &  3.7 & 1.47 & 26.3 & 26.6 & 1.10 & THIN1 \\
    RXCJ0020.7-2542	& 0201900301 & 0.1424 &  5.7 & 1.78 & 14.8 & 15.4 & 1.02 & THIN1 \\
    RXCJ0049.4-2931	& 0201900401 & 0.1080 &  3.3 & 1.37 & 19.2 & 18.8 & 1.28 & THIN1 \\
    RXCJ0547.6-3152	& 0201900901 & 0.1483 &  6.7 & 1.92 & 23.3 & 24.0 & 1.12 & THIN1 \\
    RXCJ0605.8-3518	& 0201901001 & 0.1410 &  4.9 & 1.65 & 18.0 & 24.1 & 1.07 & THIN1 \\
    RXCJ0645.4-5413	& 0201901201 & 0.1670 &  7.1 & 1.95 & 10.9 & 10.9 & 1.11 & THIN1 \\
    RXCJ1044.5-0704	& 0201901501 & 0.1323 &  3.9 & 1.47 & 25.7 & 25.9 & 1.03 & THIN1 \\
    RXCJ1141.4-1216	& 0201901601 & 0.1195 &  3.8 & 1.46 & 28.4 & 28.6 & 1.03 & THIN1 \\
    RXCJ1516.3+0005	& 0201902001 & 0.1183 &  5.3 & 1.73 & 26.7 & 26.6 & 1.13 & THIN1 \\
    RXCJ1516.5-0056	& 0201902101 & 0.1150 &  3.8 & 1.46 & 30.0 & 30.0 & 1.08 & THIN1 \\
    RXCJ2014.8-2430	& 0201902201 & 0.1612 &  7.1 & 1.96 & 23.0 & 23.4 & 1.05 & THIN1 \\
    RXCJ2048.1-1750	& 0201902401 & 0.1470 &  5.6 & 1.75 & 24.6 & 25.3 & 1.07 & THIN1 \\
    RXCJ2149.1-3041	& 0201902601 & 0.1179 &  3.3 & 1.37 & 25.1 & 25.5 & 1.11 & THIN1 \\
    RXCJ2218.6-3853	& 0201903001 & 0.1379 &  6.4 & 1.88 & 20.2 & 21.4 & 1.11 & THIN1 \\
    RXCJ2234.5-3744	& 0201903101 & 0.1529 &  8.6 & 2.17 & 18.9 & 19.3 & 1.31 & THIN1 \\
    RXCJ0645.4-5413	& 0201903401 & 0.1670 &  8.5 & 2.13 & 11.5 & 12.1 & 1.51 & THIN1 \\
    RXCJ0958.3-1103	& 0201903501 & 0.1527 &  6.1 & 1.83 &  8.3 &  9.4 & 1.16 & THIN1 \\
    RXCJ0303.8-7752	& 0205330101 & 0.2742 &  7.5 & 1.89 & 11.7 & 11.5 & 1.18 & THIN1 \\
    RXCJ0516.7-5430	& 0205330301 & 0.2952 &  7.5 & 1.87 & 11.4 & 11.7 & 1.19 & THIN1 \\
    \hline
  \end{tabular}
  \begin{list}{}{}
    \item[Notes:] $\mathrm{^a}$ redshift taken from the NASA
    Extragalactic Database; $\mathrm{^b}$ mean temperature in keV
    derived from our analysis; $\mathrm{^c}$ scale radius in Mpc
    derived from our analysis; $\mathrm{^d}$ MOS1 and MOS2 good
    exposure time in ks; $\mathrm{^e}$ intensity of residual soft
    protons (see~Eq.~\ref{eq: RSB}); $^*$ excluded due to high
    residual soft proton contamination.
    \end{list}
\end{table*}
In Table~\ref{tab: sample} we list the 48 observations that
survived our selection criteria and report cluster physical
properties.
The redshift value (from optical measurements) is taken from
the NASA Extragalactic
Database\footnote{http://nedwww.ipac.caltech.edu};
$\mathrm{k}T_\mathrm{M}$ and $R_{180}$ are derived from our
analysis (see Sect.~\ref{sec: profiles}).
\begin{figure}
  \centering
  \resizebox{80mm}{!}{\includegraphics[angle=0]{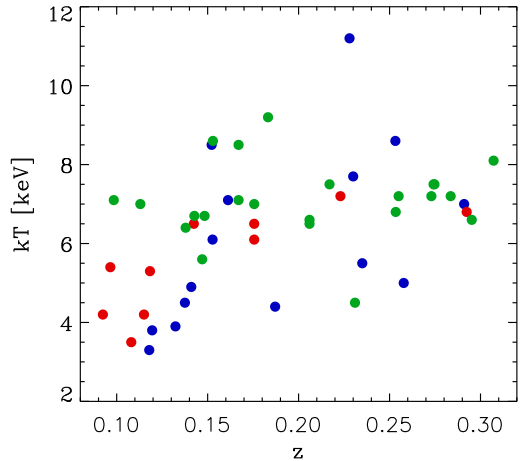}} \\
  \caption{Distribution of selected clusters in the
  redshift-temperature space. We distinguish cool core
  (blue), non cool core (red) and uncertain (green) clusters,
  as defined in Sect.~\ref{sec: mean prof}.
  There is no evidence of selection effects, except for a weak
  positive correlation between redshift and temperature.}
  \label{fig: correl}
\end{figure}
In Fig.~\ref{fig: correl} we report the cluster distribution
in the redshift-temperature space.
The only selection effect we detect is the paucity of cool
($\mathrm{k}T_\mathrm{M} \lesssim 5$~keV) clusters at high
($z > 0.2$) redshift.
Observations are performed with THIN1 and MEDIUM filters,
as reported in Table~\ref{tab: sample}.


\section{Data analysis}  \label{sec: analysis}
The preparation of spectra comprises the following major steps:
\begin{itemize}
   \item preliminary data processing;
   \item good time interval (GTI) filtering to exclude periods of high soft
         proton flux;
   \item filtering according to pattern and flag criteria;
   \item excision of brightest point-like sources;
   \item calculation of the ``IN over OUT'' ratio;
   \item extraction of spectra in concentric rings.
\end{itemize}
The spectral analysis is structured as follows:
\begin{itemize}
   \item estimate of background parameters from a peripheral ring of the
         field of view;
   \item spectral fitting using the Cash statistic and modeling the
         background, rather than subtracting it, as commonly done;
   \item production of surface brightness, temperature, and metallicity
         profiles.
\end{itemize}
All these points are described in detail in the following subsections.

In our analysis we use only EPIC-MOS data, because a robust characterization
of EPIC-pn background has not been possible, mainly due to the small regions
outside the field of view and to the non-negligible fraction of out of time
events (for further details, see Appendix~\ref{sec: blankfield}).
Moreover, the EPIC-pn background is less stable than the EPIC-MOS one,
especially below 2~keV.


\subsection{Spectra preparation} \label{sec: sp prep}

\subsubsection{Preliminary data preparation} \label{sec: prelim}
Observation data files (ODF) are retrieved from the \emph{XMM-Newton}
archive and processed in a standard way with the Science Analysis System
(SAS) v6.1.

The soft proton cleaning is performed using a double filtering process.
We extract a light curve in 100 second bins in the 10-12~keV energy band
by excluding the central CCD, apply a threshold of 0.20 cts s$^{-1}$,
produce a GTI file and generate the filtered event file accordingly.
This first step allows to eliminate most flares, however softer flares may
exist such that their contribution above 10~keV is negligible.
We then extract a light curve in the 2-5~keV band and fit the histogram
obtained from this curve with a Gaussian distribution.
Since most flares have been rejected in the previous step, the fit is
usually very good.
We calculate the mean count rate, $\mu$, and the standard deviation,
$\sigma$, apply a threshold of $\mu+3\sigma$ to the distribution, and
generate the filtered event file.

After soft proton cleaning, we filter the event file according to
\verb|PATTERN| and \verb|FLAG| criteria (namely \verb|PATTERN|$\leq$12 and
\verb|FLAG|==0).
\begin{figure}
  \centering
  \begin{tabular}{cc}
    \hspace{-7mm}
    \resizebox{48mm}{!}{\includegraphics{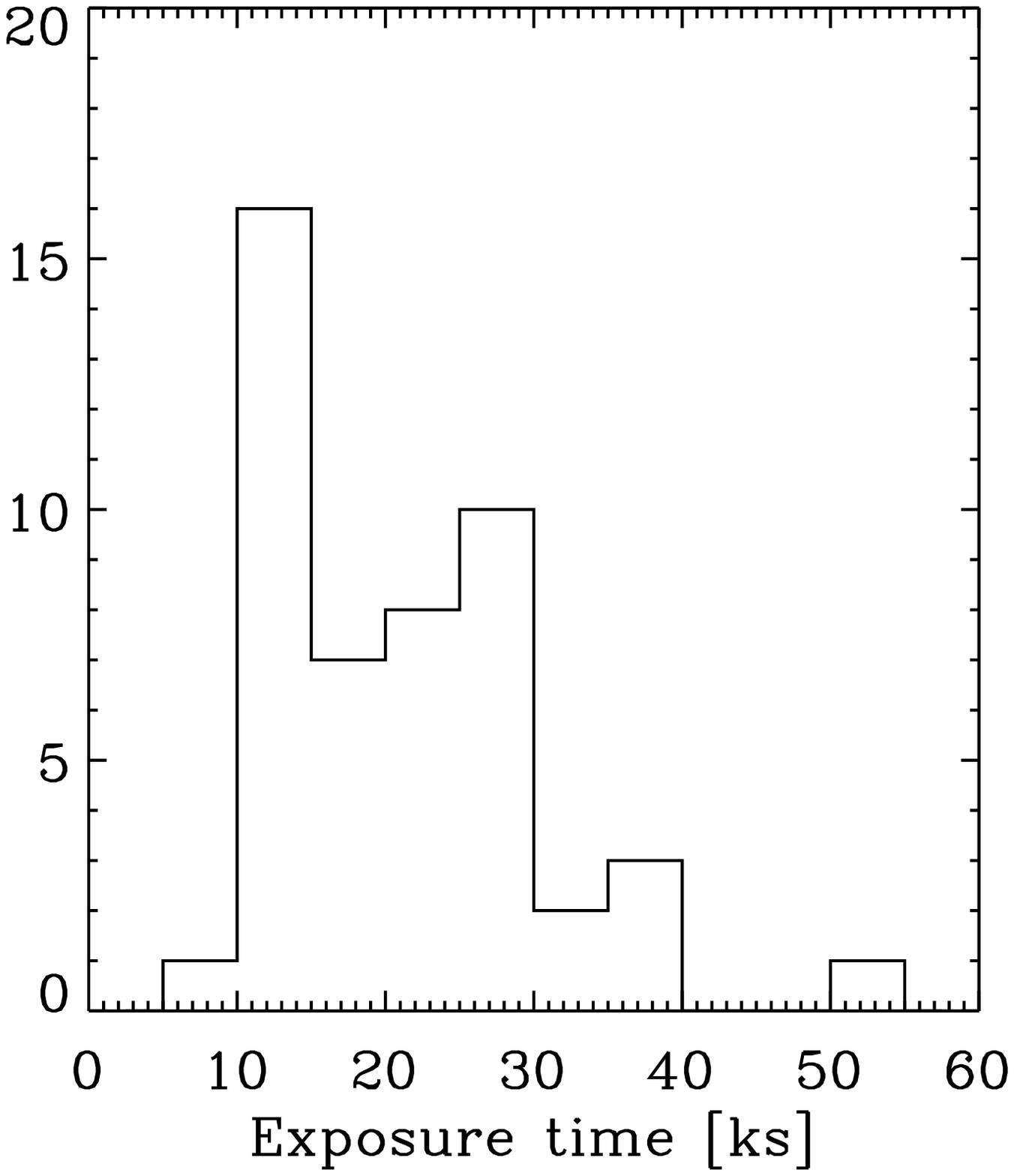}} &
    \hspace{-7mm}
    \resizebox{48mm}{!}{\includegraphics{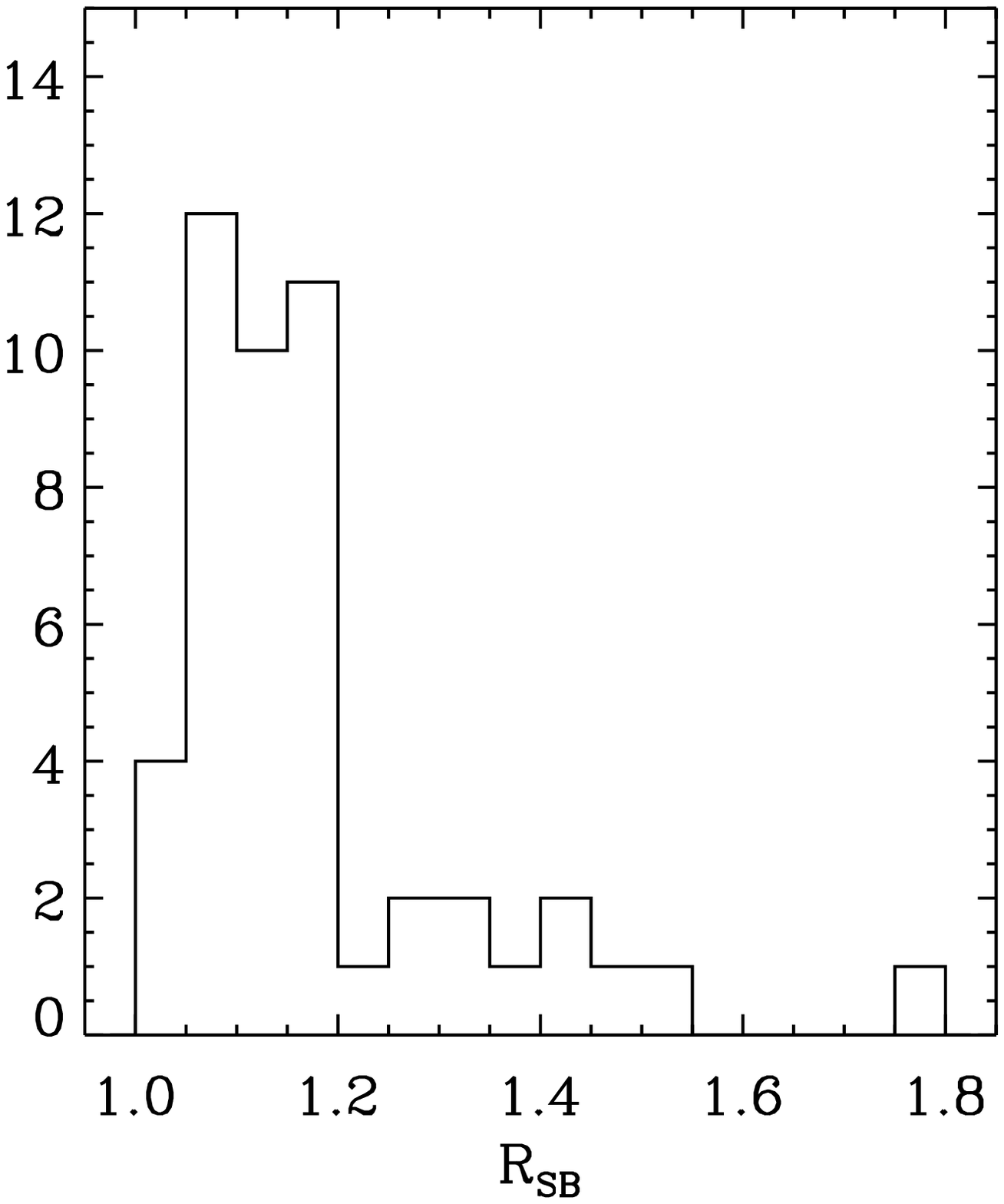}} \\
  \end{tabular}
  \caption{Histograms of the frequency distribution for averaged
  MOS exposure time (left panel) and $R_\mathrm{SB}$ (right panel)
  values.}
  \label{fig: histo}
\end{figure}
In Table~\ref{tab: sample} we report the good exposure time after the soft
proton cleaning; as mentioned in Sect.~\ref{sec: sample}, we exclude
observations for which the total (MOS1+MOS2) good exposure time is less
than 16~ks.
In the left panel of Fig.~\ref{fig: histo} we report the histogram of the
frequency distribution for observation exposure times.

When fitting spectra in the 0.7-10.0~keV band (see Sect.~\ref{sec: sp ana}),
we also exclude the ``bright'' CCDs, i.e. CCD-4 and CCD-5 for MOS1 and CCD-2
and CCD-5 for MOS2 (see Appendix~\ref{sec: closed} for the discussion).

Brightest point-like sources are detected, using a procedure based on the
SAS task \verb+edetect_chain+ and excluded from the event file.
We estimate a flux limit for excluded sources in the order of
$10^{-13}$~erg~cm$^{-2}$~s$^{-1}$; after the source excision, the cosmic
variance of the X-ray background on the entire field of view is
$\approx$~20\%.

\subsubsection{Quiescent soft proton contamination} \label{sec: INOUT}
A quiescent soft proton (QSP) component can survive the double filtering
process (see Sect.~\ref{sec: prelim}).
To quantify the amount of this component, we make use of the ``IN over OUT''
diagnostic\footnote{A public script is available at
http://xmm.vilspa.esa.es/external/\\
xmm\_sw\_cal/background/epic\_scripts.shtml} \citep{deluca04}.
We measure the surface brightness, SB$_\mathrm{IN}$, in an outer region of
the field of view, where the cluster emission is negligible, and compare it
to the surface brightness, SB$_\mathrm{OUT}$, calculated outside the field
of view in the same energy range (i.e. 6-12~keV).
Since soft protons are channeled by the telescope mirrors inside the field
of view and the cosmic ray induced background covers the whole detector,
the ratio
\begin{equation}
R_\mathrm{SB}=\frac{S\!B_\mathrm{IN}}{S\!B_\mathrm{OUT}}
\label{eq: RSB}
\end{equation}
is a good indicator of the intensity of residual soft protons and is
used for background modeling (see Sect.~\ref{sec: int rings} and
Appendix~\ref{sec: blankfield}).
In Table~\ref{tab: sample} we report the values of $R_\mathrm{SB}$ for each
observation; they roughly span the range between 1.0 (negligible
contamination) and 1.5 (high contamination).
The typical uncertainty in measuring $R_\mathrm{SB}$ is a few percent.
In the right panel of Fig.~\ref{fig: histo} we report the frequency
distribution for $R_\mathrm{SB}$ values.
Since the observation 0084230401 of Abell~267 is extremely polluted by QSP
($R_\mathrm{SB}=1.8$), we exclude it from the sample.

\subsubsection{Spectra accumulation} \label{sec: spec accum}
The cluster emission is divided in 10 concentric rings (namely
0$^\prime$-0.5$^\prime$, 0.5$^\prime$-1$^\prime$, 1$^\prime$-1.5$^\prime$,
1.5$^\prime$-2$^\prime$, 2$^\prime$-2.75$^\prime$,
2.75$^\prime$-3.5$^\prime$, 3.5$^\prime$-4.5$^\prime$,
4.5$^\prime$-6$^\prime$, 6$^\prime$-8$^\prime$, and
10$^\prime$-12$^\prime$).
The center of the rings is determined by surface brightness isocontours at
large radii and is not necessarily coincident with the X-ray emission peak.
We prefer that azimuthal symmetry be preserved at large radii, where we are
interested in characterizing profiles, at the expense of central regions.

For each instrument (i.e MOS1 and MOS2) and each ring, we accumulate a
spectrum and generate an effective area (ARF); for each observation we
generate one redistribution function (RMF) for MOS1 and one for MOS2.
We perform a minimal grouping to avoid channels with no counts, as required
by the Cash statistic.


\subsection{Spectral analysis} \label{sec: sp ana}
Spectral fitting is performed within the XSPEC v11.3
package\footnote{http://heasarc.nasa.gov/docs/xanadu/xspec/xspec11/index.html}.
The choice of the energy band for the spectral fitting is not trivial.
We fit spectra in the 0.7-10.0~keV and in the 2.0-10.0~keV energy bands, by
using the Cash statistic, with an absorbed thermal plus background model.
The high energy band has the advantage of requiring a simplified background
model (see Appendices~\ref{sec: closed} and \ref{sec: blankfield}); however,
the bulk of source counts is excluded and the statistical quality of the
measurement is substantially reduced.
Due to the paucity of source counts, there is a strong degeneracy between
source temperature and normalization, and the temperature is systematically
underestimated; therefore, when using the 2.0-10.0~keV band, an ``a
posteriori'' correction is required \citep{leccardi07}.
On the contrary, in the 0.7-10.0~keV band, the statistical quality of the
data is good, but the background model is more complicated and background
components are less stable and affected by strong degeneracy (see
Appendices~\ref{sec: closed} and \ref{sec: blankfield}).
We exclude the band below 0.7~keV because the shape of the internal
background is very complicated and variable with time and because the source
counts reach their maximum at $\approx$~1~keV.
Hereafter, all considerations are valid for both energy bands, unless
otherwise stated.

In conditions of poor statistics (i.e. few counts/bin) and high background,
the Cash statistic \citep{cash79} is more suitable than the $\chi^2$ with
reasonable channel grouping \citep{leccardi07}.
The Cash statistic requires the number of counts in each channel to be
greater than zero \citep{cash79}; thus, the background cannot be subtracted.
In our case the total background model is the sum of many components,
each one characterized by peculiar temporal, spectral, and spatial
variations (see Appendix~\ref{sec: blankfield}); when subtracting the
background, the information on single components is lost.
Conversely, background modeling allows to preserve the information and
to manage all components appropriately.
Moreover, we recall that the background modeling does not require strong
channel grouping, error propagation, or renormalization factors.

\subsubsection{Estimate of background parameters} \label{sec: ext ring}
To model the background, a careful characterization of all its components is
mandatory.
Ideally, one would like to estimate background parameters in the same region
and at the same time as the source.
Since this is not possible, we estimate background parameters in the external
10$^\prime$-12$^\prime$ ring and rescale them in the inner rings, by making
reasonable assumptions on their spatial distribution tested by analyzing
blank field observations (see Appendix~\ref{sec: blankfield}).
The 10$^\prime$-12$^\prime$ ring often contains a weak cluster emission
that, if neglected, may cause a systematic underestimate of temperature
and normalization in the inner rings (see Sect.~\ref{sec: pri ext ring}).
In this ring the spectral components in the 0.7-10.0~keV band are: 
\begin{itemize}
   \item the thermal emission from the cluster (GCL),
   \item the emission from the Galaxy Halo (HALO),
   \item the cosmic X-ray background (CXB),
   \item the quiescent soft protons (QSP),
   \item the cosmic ray induced continuum (NXB),
   \item the fluorescence emission lines;
\end{itemize}
the HALO component is negligible when considering the 2.0-10.0~keV range.
The model is the same used when analyzing blank field observations (see
Appendix~\ref{sec: blankfield} for further details) plus a thermal component
for the GCL.

We fixed most parameters (namely all except for the normalization of
HALO, CXB, NXB, and fluorescence lines) to reduce the degeneracy due to the
presence of different components with similar spectral shapes.
All cluster parameters are fixed: the temperature, $\mathrm{k}T$, and the
normalization, $N_\mathrm{S}$, are extrapolated from the final profiles
through an iterative procedure; the metallicity, $Z$, is fixed to 0.2 solar
(the solar abundances are taken from \cite{anders89}) and the redshift, $z$,
is fixed to the optical value.
The QSP normalization, $N_\mathrm{QSP}$, is calculated from $R_\mathrm{SB}$
(see Appendix~\ref{sec: blankfield}) and fixed.
Minor discrepancies in shape or normalization with respect to the real QSP
spectrum are possible; the model accounts for them by slightly changing the
normalization of other components, i.e. $N_\mathrm{HALO}$, $N_\mathrm{CXB}$,
and $N_\mathrm{NXB}$ (for the discussion of the systematic effects related
to QSP see Sects.~\ref{sec: pri QSP} and \ref{sec: post INOUT}).

Summarizing, in the 10$^\prime$-12$^\prime$ ring we determine the range of
variability, [$N_\mathrm{min}$,$N_\mathrm{max}$], (i.e. the best fit value
$\pm$1$\sigma$ uncertainty) for the normalization of the main background
components, i.e. $N_\mathrm{HALO}$, $N_\mathrm{CXB}$, and $N_\mathrm{NXB}$.
Once properly rescaled, this information allows us to constrain background
parameters in the inner rings.

\subsubsection{Spectral fit in concentric rings} \label{sec: int rings}
\begin{figure*}
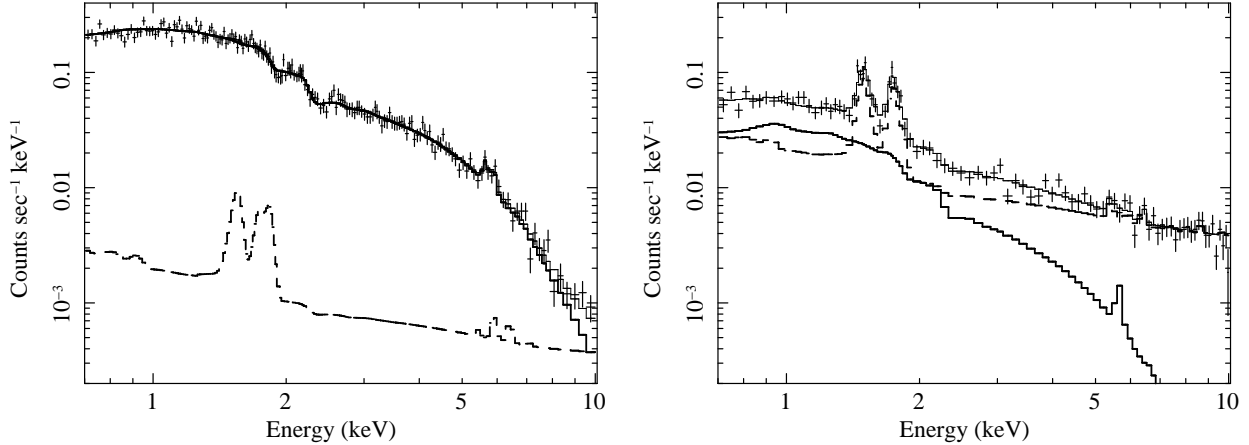

  \centering
  \begin{tabular}{cc}
    \resizebox{80mm}{!}{\includegraphics[angle=270]{reg_1_15.ps}} &
    \resizebox{80mm}{!}{\includegraphics[angle=270]{reg_45_6.ps}} \\
  \end{tabular}
  \caption{Spectra and best fit models for the 1$^\prime$-1.5$^\prime$
  (left) and the 4.5$^\prime$-6$^\prime$ (right) rings of Abell~1689. The
  solid thick and the dotted thick lines represent respectively the thermal
  and the total background model. The solid thin line represents the total
  (i.e. thermal + background) model. In the inner ring source counts
  dominate background ones, in the outer the opposite is true.}
  \label{fig: cfr spec}
\end{figure*}

We fit spectra in internal rings with the same model adopted in the
10$^\prime$-12$^\prime$ ring case (see Sect.~\ref{sec: ext ring}).
In Fig.~\ref{fig: cfr spec} we compare spectra and best fit models for two
different regions of the same cluster; in the inner ring
(1$^\prime$-1.5$^\prime$) source counts dominate, while in the outer ring
(4.5$^\prime$-6$^\prime$) background counts dominate.

The equivalent hydrogen column density along the line of sight,
$N_\mathrm{H}$, is fixed to the 21 cm measurement \citep{dickey90}.
Since clusters in our sample are at high galactic latitude
($|b| > 20^\circ$), the $N_\mathrm{H}$ is $< 10^{21}$~cm$^{-2}$
and the absorption effect is negligible above 1~keV.
We always leave the temperature, $\mathrm{k}T$, and the normalization,
$N_\mathrm{S}$, free to vary; the metallicity is free below
$\approx$~0.4~$R_{180}$ and fixed to 0.2 solar beyond; the redshift is
allowed to vary between $\pm$7\% of the optical measurement in the two
innermost rings and, in the other rings, is fixed to the average value
of the first two rings.

$N_\mathrm{HALO}$, $N_\mathrm{CXB}$, and $N_\mathrm{NXB}$ for the inner
rings are obtained by rescaling the best-fit values in the
10$^\prime$-12$^\prime$ ring (see Sect.~\ref{sec: ext ring}) by the area
ratio and the correction factor, $K(r)$, obtained from blank field
observations (see Table~\ref{tab: corr factor} in
Appendix~\ref{sec: blankfield}):
\begin{equation}
N^\mathrm{int}=N^\mathrm{ext}\times\frac{Area^\mathrm{int}}
{Area^\mathrm{ext}} \times K(r)\;,
\label{eq: rescale}
\end{equation}
for NXB $K=1$ for all rings.
$N^\mathrm{int}_\mathrm{HALO}$, $N^\mathrm{int}_\mathrm{CXB}$, and
$N^\mathrm{int}_\mathrm{NXB}$ are free to vary within a certain range: the lower
(upper) limit of this range is derived by rescaling the best-fit value minus
(plus) the 1$\sigma$-error calculated in the 10$^\prime$-12$^\prime$ ring.
The local background should have a variation length scale of some degrees
\citep{snowden97}; conversely, $N_\mathrm{CXB}$ may have large (i.e.
20-100\%) variations between different rings due to the cosmic variance.
However, extensive simulations show that these statistical fluctuations do
not introduce systematics in the temperature measurement, when averaging
on a large sample (see Sect.~\ref{sec: pri CV}).
$N^\mathrm{int}_\mathrm{QSP}$ is obtained by rescaling the value adopted
in the 10$^\prime$-12$^\prime$ ring by the area ratio and by the QSP
vignetting profile \citep{kuntz06}; $N^\mathrm{int}_\mathrm{QSP}$ is fixed
for all rings.
Normalizations of instrumental fluorescence emission lines are left free
to vary within a limited range determined from the analysis of closed
observations and have an almost negligible impact on our measurements.

For each ring, when using the 0.7-10.0~keV energy band, we determine
$\mathrm{k}T$, $Z$, and $N_\mathrm{S}$ best fit values and one sigma
uncertainties for each MOS and calculate the weighted average.
Conversely, when using the 2.0-10.0~keV band, we combine temperature
measurements from different instruments as described in our previous
paper \citep{leccardi07}, to correct for the bias which affects the
temperature estimator.
In the 0.7-10.0~keV band there are much more source counts, the
temperature estimator is much less biased and the weighted average
returns a slightly ($\approx$~3\% in an outer ring) biased value (see
the $F=1.0$ case in Sect.~\ref{sec: pri CV}).

Finally, we produce surface brightness (i.e. normalization over area),
temperature, and metallicity profiles for each cluster.


\section{The temperature profiles} \label{sec: profiles}
Clusters in our sample have different temperatures and redshifts, therefore
it is not trivial to identify one (or more) parameters that indicate the
last ring where our temperature measurement is reliable.
We define an indicator, $I$, as the source-to-background count rate ratio
calculated in the energy band used for the spectral fitting.
For each observation we calculate $I$ for each ring: the higher is $I$, the
more important is the source contribution, the more reliable is our
measurement in this particular ring.
$I$ is affected by an intrinsic bias, i.e. upward statistical fluctuations
of the temperature are associated to higher $I$ (because of the difference
in spectral shape between source and background models); therefore, near to
a threshold, the mean temperature results slightly overestimated.
This systematic is almost negligible when considering the whole sample,
but it may appear when analyzing a small number of objects.
We note that, although present, this effect does not affect results obtained
when dividing the whole sample in subsamples (e.g. Sects.~\ref{sec: post INOUT}
and \ref{sec: evolution}).

\begin{figure}
  \centering
  \resizebox{80mm}{!}{\includegraphics[angle=0]{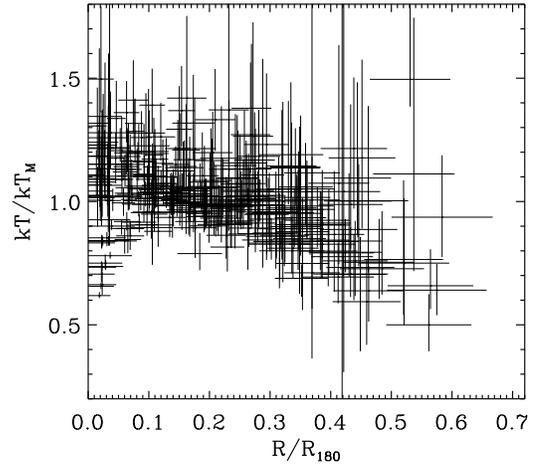}} \\
  \caption{Radial temperature profiles for all clusters in our sample
  rescaled by $R_{180}$ and $\mathrm{k}T_\mathrm{M}$.}
  \label{fig: all prof}
\end{figure}
In Fig.~\ref{fig: all prof} we show the radial temperature profiles for all
clusters of our sample by setting a lower limit $I_0=0.6$; spectra are fitted
in the 0.7-10.0~keV band.
Each profile is rescaled by the cluster mean temperature,
$\mathrm{k}T_\mathrm{M}$, computed by fitting the profile with a constant
after the exclusion of the core region (i.e. for $R > 0.1 \; R_{180}$).
The radius is rescaled by $R_{180}$, i.e. the radius encompassing a
spherical density contrast of 180 with respect to the critical density.
We compute $R_{180}$ from the mean temperature and the redshift
\citep{arnaud05}:
\begin{equation}
\label{eq: r180}
R_{180}=1780\,\left(\frac{\mathrm{k}T_\mathrm{M}}{5\,\mathrm{keV}}\right)^{1/2}
h(z)^{-1} \; \mathrm{kpc} ,
\end{equation}
where $h(z)=(\Omega_\mathrm{m}(1+z)^3+\Omega_\Lambda)^{1/2}$.
$R_{180}$ is a good approximation to the virial radius in an
Einstein-De~Sitter universe and has been largely used to rescale cluster
radial properties \citep{grandi02,vikh05}.
We then choose 180 as over-density for comparing our results with previous works
(see Sect~\ref{sec: comp obs}), even if in the current adopted cosmology
the virial radius encloses a spherical density contrast of $\approx$~100
\citep{eke98}.

The profiles show a clear decline beyond $\approx$~0.2~$R_{180}$ and our
measurements extend out to $\approx$~0.6~$R_{180}$.
The large scatter of values is mostly of statistical origin, however
a maximum likelihood test shows that, when excluding the region below
0.2~$R_{180}$, our profiles are characterized by a 6\% intrinsic
dispersion, which is comparable with our systematics
(see Sect.~\ref{sec: syst budget}), therefore the existence of a universal
cluster temperature profile is still an open issue.
The scatter in the inner region is mostly due to the presence of
both cool core and non cool core clusters, but also to our
choice of preserving the azimuthal symmetry at large radii
(see Sect.~\ref{sec: spec accum}).
\begin{figure}
  \centering
  \resizebox{80mm}{!}{\includegraphics[angle=0]{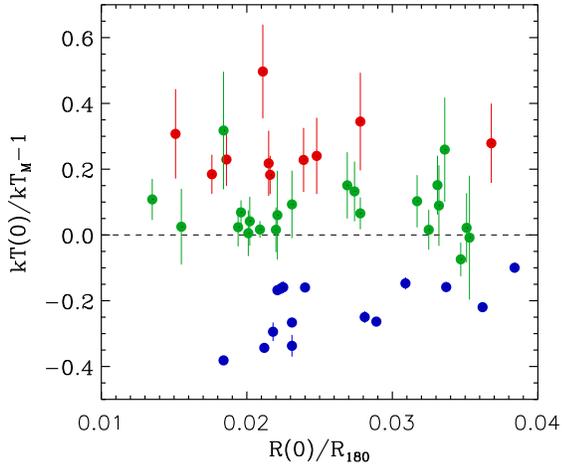}} \\
  \caption{Temperature vs. radius for the innermost ring respectively
  scaled by   $\mathrm{k}T_\mathrm{M}$ and $R_{180}$. Clusters for
  which the temperature is significantly (at least $3 \sigma$) lower
  than $\mathrm{k}T_\mathrm{M}$ are defined as cool cores (blue circles);
  those for which the temperature profile does not significantly
  (at least $2 \sigma$) decrease are defined as non cool cores
  (red circles); other clusters, whose membership is not clearly
  determined, are classified as uncertain (green circles).
  When considering $z>0.2$ clusters, which fill the right-side of
  the panel, we expect smaller gradients due to the lower spatial
  resolution.}
  \label{fig: def CC}
\end{figure}
In Fig.~\ref{fig: def CC} we report temperature and radius of the
innermost ring scaled by $\mathrm{k}T_\mathrm{M}$ and $R_{180}$
for all clusters.
We define cool core (hereafter CC) clusters, those for which the
temperature is significantly (at least $3 \sigma$) lower than
$\mathrm{k}T_\mathrm{M}$, non cool core (hereafter NCC) clusters,
those for which the temperature profile does not significantly (at
least $2 \sigma$) decrease, and uncertain (hereafter UNC) clusters,
those for which the membership is not clearly determined.

It is worth noting that the error bars are usually strongly asymmetric,
i.e. the upper bar is larger than the lower; moreover, the higher the
temperature, the larger the error bars.
The reason is that most of the information on the temperature is located
around the energy of the exponential cut-off; due to the spectral shapes
of source and background components and to the sharp decrease of the
effective areas at high energies, the source-to-background count rate
ratio strongly depends on the energy band (see for example
Fig.~\ref{fig: cfr spec}), i.e. the higher the cut-off energy, the lower
the source-to-background ratio, the larger the uncertainties.

\begin{figure}
  \centering
  \resizebox{80mm}{!}{\includegraphics[angle=0]{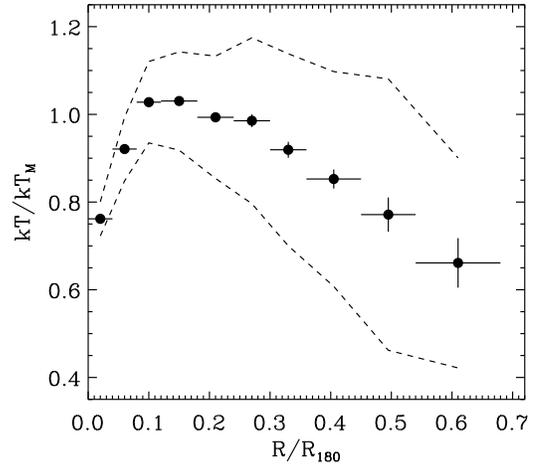}} \\
  \caption{Mean radial temperature profile rescaled by $R_{180}$
  and $\mathrm{k}T_\mathrm{M}$. The dotted lines show the one-sigma
  scatter of the values around the average.}
  \label{fig: mean prof}
\end{figure}
In Fig.~\ref{fig: mean prof} we report the weighted average and the
scatter of all profiles shown in Fig.~\ref{fig: all prof}.
The mean profile shows more clearly the decline beyond 0.2~$R_{180}$.
The temperature also decreases toward the center because of the
presence of cool core clusters.


\section{Evaluation of systematic effects} \label{sec: syst}
We carefully check our results, searching for possible systematic effects.
Prior to the analysis, we make use of extensive simulations to quantify
the impact of different spectral components on a simulated temperature
profile (``a priori'' tests).
After the analysis, we investigate how the measured temperature profile
changes, when choosing different key parameters (``a posteriori'' tests).

\subsection{``A priori'' tests} \label{sec: pri}
We perform simulations that reproduce as closely as possible our analysis
procedure.
We consider two rings: the external 10$^\prime$-12$^\prime$,
$R_\mathrm{ext}$, where we estimate background parameters, and the
4.5$^\prime$-6$^\prime$, $R_\mathrm{int}$, where we measure the
temperature.
The exposure time for each spectrum is always 20~ks i.e. a representative
value for our sample (see Fig.~\ref{fig: histo}).
We use the Abell~1689 EPIC-MOS1 observation as a guideline, for producing
RMF and ARF, and for choosing typical input parameters.
The simulation procedure is structured as follows:
\begin{itemize}
   \item choice of reasonable input parameters,
   \item generation of 300 spectra in $R_\mathrm{ext}$,
   \item generation of 500 spectra in $R_\mathrm{int}$,
   \item estimate of background parameters in $R_\mathrm{ext}$,
   \item rescaling background parameters and fitting spectra in $R_\mathrm{int}$.
\end{itemize}
Simulation details are described in each subsection.
We test the effect of the cosmic variance (see Sect.~\ref{sec: pri CV}),
of an inaccurate estimate of the cluster emission in $R_\mathrm{ext}$ (see
Sect.~\ref{sec: pri ext ring}), and of the QSP component (see
Sect.~\ref{sec: pri QSP}).
All results are obtained by fitting spectra in the 0.7-10.0~keV band.
We have also conducted a similar analysis for the 2.0-10.0~keV band and
have found that the systematics for the two bands are of the same order of
magnitude.
We recall however that the hard band is characterized by worst statistics,
therefore in this case systematic errors are masked by statistical ones and
have a smaller impact on the final measurement.

\subsubsection{The cosmic variance} \label{sec: pri CV}
We employ a simulation to quantify the effect of the cosmic variance
on temperature and normalization measurements.
In this simulation we neglect the soft proton contribution; the
background components are the HALO, the CXB, and the NXB and they are
modeled as for MOS1 in Appendix~\ref{sec: blankfield}.
In $R_\mathrm{ext}$ there are only background components, while in
$R_\mathrm{int}$ there is also the thermal source.
Normalization\footnote{Normalization values are always reported in
XSPEC units} input values in $R_\mathrm{ext}$ are:
$N^\mathrm{ext}_\mathrm{HALO}=1.6\times10^{-4}$,
$N^\mathrm{ext}_\mathrm{CXB}=5.0\times10^{-2}$, and
$N^\mathrm{ext}_\mathrm{NXB}=1.0\times10^{-2}$; input values in
$R_\mathrm{int}$ are obtained by rescaling the values in $R_\mathrm{ext}$
by the area ratio (i.e. as in Eq.~\ref{eq: rescale} with $K(r)=1.0$).
$N_\mathrm{CXB}$ is also multiplied by a factor, $F$, that simulates the
fluctuation due to the cosmic variance between $R_\mathrm{int}$ and
$R_\mathrm{ext}$; after the excision of brightest point-like sources
(see Sect.~\ref{sec: prelim}), 1$\sigma$ fluctuations are expected to
be $\approx$~30\%.
We then consider 3 cases: a null ($F=1.0$), a positive ($F=1.3$), and a
negative ($F=0.7$) fluctuation.
Thus, in the first case the input value for CXB in $R_\mathrm{int}$ is
equal to that rescaled by the area ratio, in the second it is 30\% higher,
and in the third 30\% lower.
Input parameters for the thermal model in $R_\mathrm{int}$ are:
$\mathrm{k}T=6$~keV, $Z=0.2\;Z_\odot$, $z=0.2$, and
$N_\mathrm{S}=7.0\times10^{-4}$.
In $R_\mathrm{ext}$, $Z$ and $z$ are fixed to the input values, while
$\mathrm{k}T$ and $N_\mathrm{S}$ are free.
For this particular choice of the parameters, the source-to-background
count rate ratio, $I$, is 1.13 (see Sect.\ref{sec: profiles}).
As explained in Sects.~\ref{sec: ext ring} and \ref{sec: int rings}, we
determine the ranges of variability for $N_\mathrm{HALO}$, $N_\mathrm{CXB}$,
and $N_\mathrm{NXB}$ and rescale them in $R_\mathrm{int}$; then we fit
spectra in the 0.7-10.0~keV band and calculate the weighted averages of
$\mathrm{k}T$ and $N_\mathrm{S}$ over the 500 simulations.

\begin{figure}
  \centering
  \resizebox{80mm}{!}{\includegraphics[angle=0]{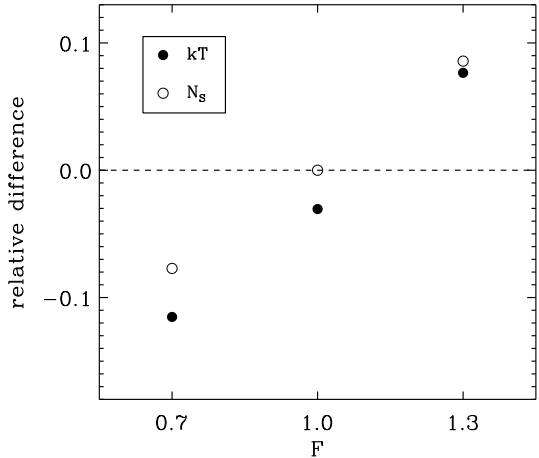}} \\
  \caption{Relative differences between measured and input
  values for the source temperature, $\mathrm{k}T$, and normalization,
  $N_\mathrm{S}$, as a function of the factor $F$, which simulates
  the fluctuation due to the cosmic variance (see text for details).
  Uncertainties are smaller than the circle size. 30\% fluctuations
  cause $\approx$~10\% variations in $\mathrm{k}T$ and $N_\mathrm{S}$.
  For a positive/negative fluctuation the measured $\mathrm{k}T$ and
  $N_\mathrm{S}$ are higher/lower than the input values.}
  \label{fig: pri CV}
\end{figure}
In Fig.~\ref{fig: pri CV} we show the relative differences between
measured and input values for the temperature, $\mathrm{k}T$
(filled circles), and the normalization, $N_\mathrm{S}$ (empty circles).
A positive fluctuation of CXB normalization (i.e. $F=1.3$) returns
higher temperature and normalization, because the excess of counts due
to the CXB is modeled by the thermal component, which is steeper than
the CXB power law.
For the $F=1.0$ case, while $N_\mathrm{S}$ returns exactly the input
value, $\mathrm{k}T$ returns a slightly ($\approx$~3\%) underestimated
value, probably due to the bias on the temperature estimator
\citep{leccardi07}.
The effect of the cosmic variance is roughly symmetric on both
$\mathrm{k}T$ and $N_\mathrm{S}$, therefore it is almost negligible
when averaging on a large sample.
We also perform simulations for our worst case, i.e. $I=0.6$ (see
Sect.~\ref{sec: profiles}), and find qualitatively the same results:
for the $F=1.0$ case, the bias on the temperature is $\approx$~8\% rather
than $\approx$~3\% and the bias on the normalization is negligible.

\subsubsection{The cluster emission in the 10$^\prime$-12$^\prime$ ring}
\label{sec: pri ext ring}
\begin{figure}
  \centering
  \resizebox{80mm}{!}{\includegraphics[angle=0]{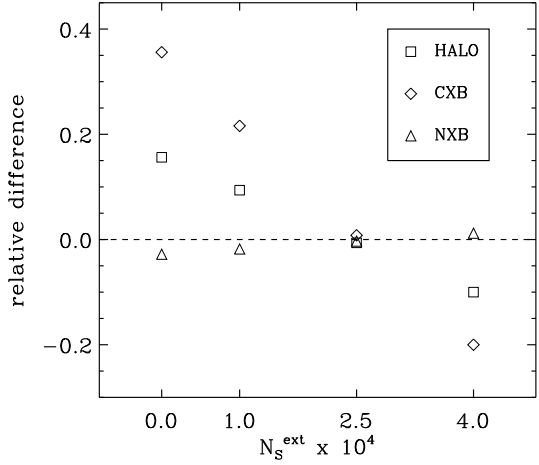}} \\
  \caption{Relative differences between measured and input
  values for the normalization of background components (namely
  $N_\mathrm{HALO}$, $N_\mathrm{CXB}$, and $N_\mathrm{NXB}$) as a
  function of the input value for cluster normalization
  in $R_\mathrm{ext}$, $N_\mathrm{S}^\mathrm{ext}$. Uncertainties
  are smaller than the symbol size. $N_\mathrm{CXB}$ shows the
  strongest (negative) correlation with $N_\mathrm{S}^\mathrm{ext}$.}
  \label{fig: pri ext1}
\end{figure}

The source contribution in the 10$^\prime$-12$^\prime$ ring, which
mainly depends on cluster redshift and emission measure, is difficult
to estimate with accuracy.
We employ a simulation to determine how an inaccurate estimate
could affect our measurement of cluster temperature, $\mathrm{k}T$,
and normalization, $N_\mathrm{S}$.
Soft protons are neglected in this case too; background components
and their input values are the same as for the $F=1.0$ case of the
cosmic variance tests (see Sect.~\ref{sec: pri CV}).
Also input parameters for the thermal model in $R_\mathrm{int}$ are
the same as in that case, instead in $R_\mathrm{ext}$ are
$\mathrm{k}T^\mathrm{ext}=4$~keV, $Z^\mathrm{ext}=0.2\;Z_\odot$,
$z^\mathrm{ext}=0.2$, and $N_\mathrm{S}^\mathrm{ext}=2.5\times10^{-4}$.
For this particular choice of the parameters, the source-to-background
count rate ratio, $I$, is 1.13 (see Sect.\ref{sec: profiles}).
When fitting spectra in $R_\mathrm{ext}$, all thermal parameters
are fixed: namely the temperature, the metallicity, and the
redshift are fixed to the input values, while for
$N_\mathrm{S}^\mathrm{ext}$ we consider 4 cases.
In the first case, we neglect the source contribution
($N_\mathrm{S}^\mathrm{ext}=0$); in the other cases, the normalization
is fixed to a value lower ($N_\mathrm{S}^\mathrm{ext}=1.0\times10^{-4}$),
equal ($N_\mathrm{S}^\mathrm{ext}=2.5\times10^{-4}$), and higher
($N_\mathrm{S}^\mathrm{ext}=4.0\times10^{-4}$) than the input value.
Normalizations of all background components (namely
$N_\mathrm{HALO}$, $N_\mathrm{CXB}$, and $N_\mathrm{NXB}$) are
free parameters.
For each case, we compute the weighted average of $N_\mathrm{HALO}$,
$N_\mathrm{CXB}$, and $N_\mathrm{NXB}$ over the 300 spectra in
$R_\mathrm{ext}$ and compare them to the input values
(see Fig.~\ref{fig: pri ext1}).
$N_\mathrm{NXB}$ and $N_\mathrm{S}^\mathrm{ext}$ are weakly correlated;
instead, $N_\mathrm{HALO}$ and, in particular, $N_\mathrm{CXB}$ show
a strong negative correlation with the input value for
$N_\mathrm{S}^\mathrm{ext}$, which depends on their spectral shapes.
Note that, if we correctly estimate $N_\mathrm{S}^\mathrm{ext}$ then
$N_\mathrm{HALO}$, $N_\mathrm{CXB}$, and $N_\mathrm{NXB}$ converge
to their input values.

\begin{figure}
  \centering
  \resizebox{80mm}{!}{\includegraphics[angle=0]{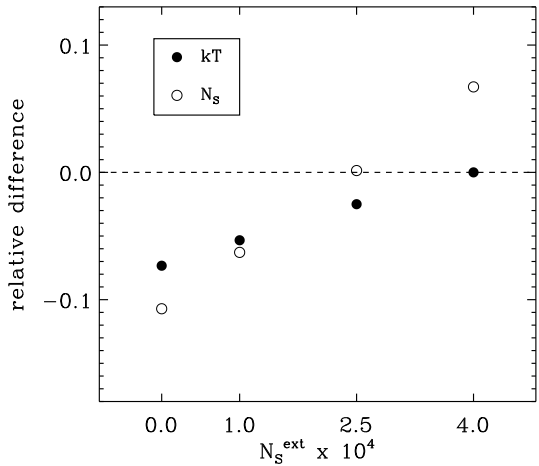}} \\
  \caption{Relative differences between measured and input
  values for the source temperature, $\mathrm{k}T$, and normalization,
  $N_\mathrm{S}$, as a function of the input value for cluster
  normalization in $R_\mathrm{ext}$, $N_\mathrm{S}^\mathrm{ext}$.
  Uncertainties are smaller than the symbol size.
  An underestimate/overestimate of $N_\mathrm{S}^\mathrm{ext}$
  causes $\mathrm{k}T$ and $N_\mathrm{S}$ to be
  underestimated/overestimated.}
  \label{fig: pri ext2}
\end{figure}
For each input value of $N_\mathrm{S}^\mathrm{ext}$ in $R_\mathrm{ext}$,
we fit spectra in $R_\mathrm{int}$ in the 0.7-10.0~keV band after
the usual rescaling of background parameters (see
Sect.~\ref{sec: int rings}), calculate the weighted averages of
the source temperature, $\mathrm{k}T$, and normalization,
$N_\mathrm{S}$, over the 500 simulations, and compare them to the
input values (see Fig.~\ref{fig: pri ext2}).
Values of $\mathrm{k}T$ and $N_\mathrm{S}$ measured in
$R_\mathrm{int}$ show a positive correlation with the value of
$N_\mathrm{S}^\mathrm{ext}$ fixed in $R_\mathrm{ext}$.
This is indeed expected because of the broad similarity in the
spectral shapes of thermal and CXB models.
In $R_\mathrm{ext}$ an overestimate of $N_\mathrm{S}^\mathrm{ext}$
implies an underestimate of $N_\mathrm{CXB}$ (see
Fig.~\ref{fig: pri ext1}); $N_\mathrm{CXB}$ is then rescaled by the
area ratio, thus is underestimated in $R_\mathrm{int}$ too; this
results in an overestimate of $\mathrm{k}T$ and $N_\mathrm{S}$ in
$R_\mathrm{int}$, as for the $F=1.3$ case of the cosmic variance
simulations (see Sect.~\ref{sec: pri CV}).
Typical uncertainties ($\approx 50\%$) on $N_\mathrm{S}^\mathrm{ext}$
cause systematic 5\% and 7\% errors on $\mathrm{k}T$ and $N_\mathrm{S}$
(see Fig.~\ref{fig: pri ext2}).
Note that, after the correction for the $\approx$~3\% bias mentioned
in Sect.~\ref{sec: pri CV}, the effect on $N_\mathrm{S}$ and
$\mathrm{k}T$ is symmetric; thus, when averaging on a large sample,
the effect on the mean profile should be almost negligible.
Note also that if we were to neglect the cluster emission in the
10$^\prime$-12$^\prime$ ring ($N_\mathrm{S}^\mathrm{ext}=0$),
we would cause a systematic underestimate of $\mathrm{k}T$ and
$N_\mathrm{S}$ in the order of 7-10\% (see Fig.~\ref{fig: pri ext2}).

In a real case we deal with a combination of fluctuations and cannot
treat each one separately, thus we employ a simulation to investigate
how fluctuations with different origins combine with each other.
We combine effects due to the cosmic variance and to an inaccurate
estimate of the cluster emission in the 10$^\prime$-12$^\prime$ ring,
by considering the $F=0.7$, $F=1.0$, and $F=1.3$ cases mentioned in
Sect.~\ref{sec: pri CV} and $N_\mathrm{S}^\mathrm{ext}=1.0\times10^{-4}$,
$N_\mathrm{S}^\mathrm{ext}=2.5\times10^{-4}$, and
$N_\mathrm{S}^\mathrm{ext}=4.0\times10^{-4}$ mentioned in this section.
The simulation procedure is the same as described before.
For the cluster normalization, we find that fluctuations combine
in a linear way and that effects are highly symmetric with respect to
the zero case ($F=1.0$ for the cosmic variance and
$N_\mathrm{S}^\mathrm{ext}=2.5\times10^{-4}$ for the cluster emission
in the 10$^\prime$-12$^\prime$ ring).
For the cluster temperature, we find again the $\approx$~3\%
bias related to the estimator; once accounted for this 3\% offset,
results are roughly similar to those found  for the normalization case.
To be more quantitative, when averaging on a large sample, the
expected systematic on the temperature measurement is
$\approx$~3\% due to the biased estimator and $\lesssim$~2\% due to
deviations from the linear regime.

\subsubsection{The QSP component} \label{sec: pri QSP}
A careful characterization of the QSP component is crucial for our
data analysis procedure.
We employ a simulation to quantify how an incorrect estimate
of the QSP contribution from the ``IN over OUT'' diagnostic, i.e.
the $R_\mathrm{SB}=1.10$ (see Sect.~\ref{sec: INOUT}) could
affect our measurements.
The spectral components and their input values are the same as
for the $F=1.0$ case of the cosmic variance simulations (see
Sect.~\ref{sec: pri CV}), plus the QSP component in both rings.
The model for QSP is the same as described in
Appendix~\ref{sec: blankfield}.
We choose two input values for $N_\mathrm{QSP}$ corresponding to
a standard ($R_\mathrm{SB}=1.10$) and a high ($R_\mathrm{SB}=1.40$)
level of QSP contamination.
For these particular choices of the parameters, the
source-to-background count rate ratio, $I$, is 1.06 for
$R_\mathrm{SB}=1.10$ and 0.77 for $R_\mathrm{SB}=1.40$
(see Sect.\ref{sec: profiles}).
For each input value we consider 2 cases: an underestimate
($R_\mathrm{SB}=1.05-1.35$) and an overestimate
($R_\mathrm{SB}=1.15-1.45$) of the correct value.
By fitting spectra in $R_\mathrm{ext}$ in the 0.7-10.0~keV
band, we determine the range of variability of $N_\mathrm{HALO}$,
$N_\mathrm{CXB}$, and $N_\mathrm{NXB}$ and rescale it in
$R_\mathrm{int}$ (see Sect.~\ref{sec: int rings}).
\begin{figure}
  \centering
  \resizebox{80mm}{!}{\includegraphics[angle=0]{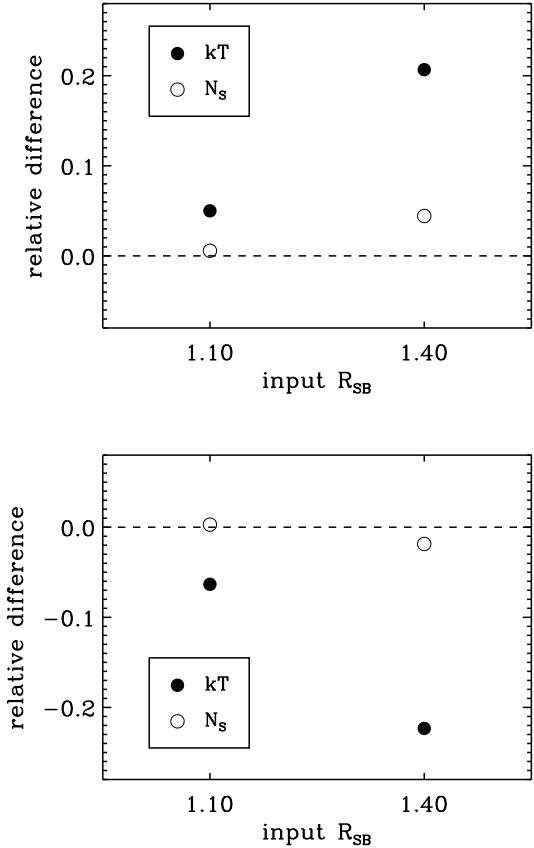}} \\
  \caption{Relative differences between measured and input
  values for the source temperature, $\mathrm{k}T$, and normalization,
  $N_\mathrm{S}$, as a function of the input value for the QSP
  contribution, $R_\mathrm{SB}$. Uncertainties are smaller than
  the circle size. Upper panel: $R_\mathrm{SB}$ is underestimated
  to 1.05 and 1.35 with respect to 1.10 and 1.40. Lower panel:
  $R_\mathrm{SB}$ is overestimated to 1.15 and 1.45. See text
  for the discussion.}
  \label{fig: pri QSP}
\end{figure}
We then fit spectra in $R_\mathrm{int}$ and compare the weighted
averages of cluster temperature, $\mathrm{k}T$, and
normalization, $N_\mathrm{S}$, to their input values
(see Fig.~\ref{fig: pri QSP}).

When considering $N_\mathrm{S}$, the relative difference between
measured and input values is $< 5\%$ for all cases and the effect
is symmetric, therefore the impact on the mean profile obtained
from a large sample should be very small.
On the contrary, $\mathrm{k}T$ strongly depends on our estimate
of the QSP component: the relative difference is $\approx$~5\% for
$R_\mathrm{SB}=1.10$ and $\approx$~20\% for $R_\mathrm{SB}=1.40$.
When overestimating $R_\mathrm{SB}$, $\mathrm{k}T$ is
underestimated, because of the broad similarity in the spectral
shapes of the two components.
In the $R_\mathrm{SB}=1.40$ case, the values corresponding to
an overestimate and an underestimate, although symmetric with
respect to zero, are characterized by different uncertainties
(errors in the first case are twice than in the second);
thus, a weighted average returns a 10\% underestimated value.

\subsection{``A posteriori'' tests}
In this subsection we investigate how the mean profile is affected
by a particular choice of key parameters, namely: the last ring for
which we measure a temperature (see Sect.~\ref{sec: post trunc}),
the energy band used for the spectral fitting (see
Sect.~\ref{sec: post bands}), and the QSP contamination (see
Sect.~\ref{sec: post INOUT}).

\subsubsection{The truncation radius} \label{sec: post trunc}
In Sect.~\ref{sec: profiles} we have introduced the indicator $I$ to
choose the last ring where our temperature measurement is reliable.
Here we produce mean temperature profiles by averaging all measurements
for which $I > I_0$, for different values of the threshold $I_0$.
\begin{figure}
  \centering
  \resizebox{80mm}{!}{\includegraphics{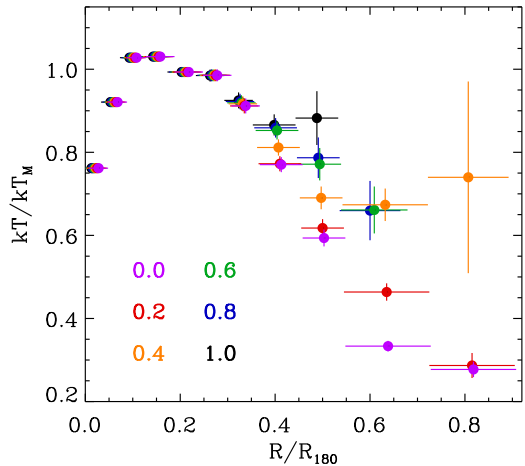}} \\
  \caption{Mean temperature profiles computed by choosing different
  values for the threshold $I_0$ (defined in Sect.~\ref{sec: profiles})
  plotted with different colors. There is a clear systematic effect:
  the smaller the threshold, the steeper the profile.
  The radii have been slightly offset in the plot for clarity.}
  \label{fig: crate ratio}
\end{figure}
In Fig.~\ref{fig: crate ratio} we report the profiles obtained in the
0.7-10.0~keV band for different choices of $I_0$ (namely 0.0,
0.2, 0.4, 0.6, 0.8, and 1.0).
As expected, the smaller is the threshold, the further the mean profile
extends.
If we focus on the points between 0.3 and 0.6 of $R_{180}$, we notice a
clear systematic effect: the smaller the threshold, the lower the
temperature.
This means that, on average, the temperature is lower in those rings
where the background is more important.
This systematic effect becomes evident where cluster emission and
background fluctuations are comparable and is probably related to small
imperfections in our background modeling and to the bias on the
temperature estimator (see Sect.~\ref{sec: pri CV}).
The imperfections of our background model becomes the dominant effect
for small values of $I$ (namely $I \lesssim 0.4$).
Thus, under a certain threshold, $I_0$, our measurements are no longer
reliable.
Fig.~\ref{fig: crate ratio} shows that $I_0=0.6$ represents a good
compromise.
Indeed, when considering the region between 0.4 and 0.5 of $R_{180}$ and
comparing the average value for $\mathrm{k}T$ obtained for a threshold
$I_0=0.6$ and for $I_0=1.0$, we find a small ($4\% \pm 3\%$) relative
difference.

\subsubsection{Fitting in different bands} \label{sec: post bands}
We have fitted spectra in two different energy bands (i.e. 0.7-10.0~keV and
2.0-10.0~keV), each one characterized by different advantages and drawbacks
(see Sect.~\ref{sec: sp ana}).
The indicator, $I$, defined in Sect.~\ref{sec: profiles} depends on the band in
which the count rate is calculated: more precisely, $I$(0.7-10.0) is roughly
1.5 times greater than $I$(2.0-10.0) for small values (i.e. $I \lesssim 2.0$).
The threshold $I_0=0.6$ in the 0.7-10.0~keV band corresponds to $I_0=0.4$
in the 2.0-10.0~keV band (see Sect.~\ref{sec: post trunc}).
\begin{figure}
  \centering
  \resizebox{80mm}{!}{\includegraphics[angle=0]{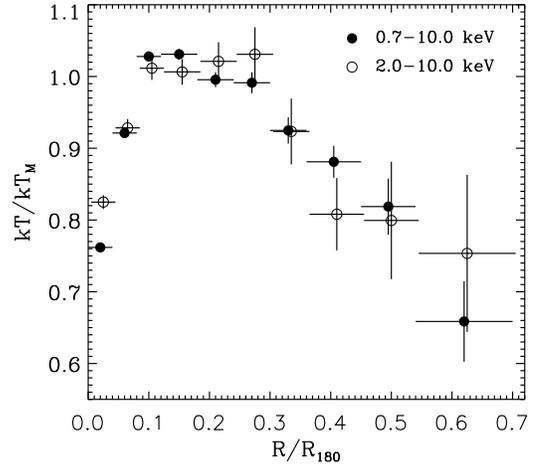}} \\
  \caption{Mean temperature profiles obtained by fitting spectra in the
  0.7-10.0~keV (filled circles) and in the 2.0-10.0~keV band (empty circles).
  The profiles are very similar, except for the innermost point. The radii
  have been slightly offset in the plot for clarity.}
  \label{fig: diff bands}
\end{figure}
In Fig.~\ref{fig: diff bands} we compare the mean temperature profile obtained
in the 0.7-10.0~keV band ($I_0=0.6$) with that obtained in the 2.0-10.0~keV
band ($I_0=0.4$).
The profiles are very similar, except for the innermost point.
The uncertainties in the 0.7-10.0 case are much smaller at all radii, even if
the total number of points (i.e. the number of rings for all cluster) is the
same; this is because the higher statistics at low energies allows to
substantially reduce the errors on single measurements.

In the most internal point a high discrepancy between the two measurements is
present, although in that region the background is negligible.
This is due to the superposition, along the line of sight, of photons emitted
by optically thin ICM with different density and temperature.
When looking at the center of cool core clusters, the line of sight intercepts
regions characterized by strong temperature gradients, therefore the accumulated
spectrum is the sum of many components at different temperatures.
In this case, the best fit value for the temperature strongly depends on
the energy band (i.e. the harder the band, the higher the temperature), because
the exclusion of the soft band implies the exclusion of most of the emission
from cooler components \citep{mazzotta04}.

\subsubsection{Contamination from QSP} \label{sec: post INOUT}
We divide clusters in our sample in four groups, according to the QSP
contamination that we estimate from $R_\mathrm{SB}$ (see
Sect.~\ref{sec: INOUT}).
\begin{figure}
  \centering
  \resizebox{80mm}{!}{\includegraphics[angle=0]{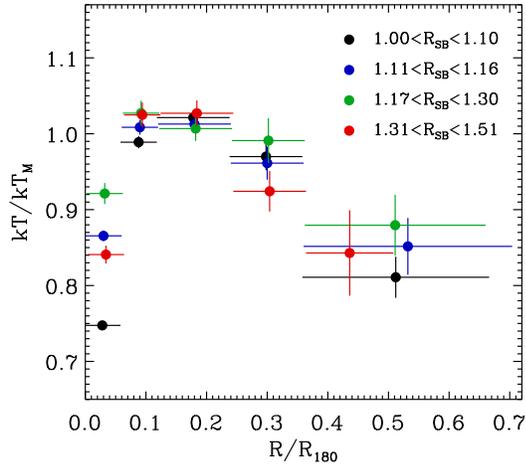}} \\
  \caption{Mean temperature profiles as a function of the QSP contamination,
  $R_\mathrm{SB}$. The four profiles are fully consistent, no correlation
  is found between the shape of the profiles and $R_\mathrm{SB}$.
  The radii have been slightly offset in the plot for clarity.}
  \label{fig: post INOUT}
\end{figure}
In Fig.~\ref{fig: post INOUT} we report the mean temperature profiles for the
four groups, by fitting spectra in the 0.7-10.0~keV band and fixing $I_0=0.6$.
When dividing clusters in subsamples, we choose larger bin sizes to
reduce the error bars.
When $R_\mathrm{SB}$ is high, our selection criterion based on the
source-to-background count rate ratio (see Sects.~\ref{sec: profiles} and
\ref{sec: post trunc}) excludes the outer rings, indeed the red profile extends
out to only 0.5~$R_{180}$.
The four profiles are fully consistent, no correlation is found between the
shape of the profiles and $R_\mathrm{SB}$.
The discrepancy in the innermost ring is due to the presence of a different
number of cool core clusters in each group.
We therefore conclude that the systematic error associated to the QSP
contamination is smaller than statistical errors ($\approx$~7\% beyond
0.4~$R_{180}$).

\subsection{A budget for systematics} \label{sec: syst budget}
In this subsection we summarize the main results for what concern systematic
errors associated to our mean profile.
We compare expected systematics computed from ``a priori'' tests with
measured systematics from ``a posteriori'' tests.

The $F=1.0$ case in Sect.~\ref{sec: pri CV} and the
$N_\mathrm{S}^\mathrm{ext}=2.5\times10^{-4}$ case in
Sect.~\ref{sec: pri ext ring} show that our analysis procedure is
affected by a 3\% to 8\% systematic underestimate of the temperature,
when analyzing the outermost rings; the bias is probably related to
the temperature estimator as described in \cite{leccardi07}.
On the contrary the normalization estimator is unbiased.
In Sects.~\ref{sec: pri CV} and \ref{sec: pri ext ring} we also found
that the effects of the cosmic variance and of an inaccurate estimate
of the cluster emission in the external ring are symmetric for both
the temperature, $\mathrm{k}T$, and the normalization, $N_\mathrm{S}$.
In Sect.~\ref{sec: pri ext ring} we found that the effects due to
fluctuations with different origins combine in a linear way and, when
averaging on a large sample, the systematic associated to the mean
profile is almost negligible for $N_\mathrm{S}$ and $\lesssim$~2\%
for $\mathrm{k}T$.
Thus, the expected systematic for $\mathrm{k}T$ is $\lesssim$~5\%.

In Sect.~\ref{sec: pri QSP} we found that, for a standard level of
contamination ($R_\mathrm{SB}=1.10$), a typical 5\% error in the estimate
of $R_\mathrm{SB}$ causes negligible effects on both measurements of
cluster temperature and normalization.
The same error causes negligible effects on $N_\mathrm{S}$ measurements
also for a high level of contamination ($R_\mathrm{SB}=1.40$).
On the contrary, effects on $\mathrm{k}T$ for $R_\mathrm{SB}=1.40$
are important: the same 5\% error causes a 10\% underestimate of
$\mathrm{k}T$, also when averaging on a large sample.
However, at the end of Sect.~\ref{sec: post INOUT} in particular from
Fig.~\ref{fig: post INOUT}, we have concluded that, when considering
the whole sample, the systematic error associated to the QSP contamination
is smaller than statistical errors ($\approx$~7\% beyond 0.4~$R_{180}$).
The difference between expected and measured systematic errors is only
apparent.
Indeed, when analyzing our sample, we average measurements that span a
wide range of values for $R_\mathrm{SB}$ and $I$; conversely, the 10\%
systematic error is expected for an unfavorable case, i.e.
$R_\mathrm{SB}=1.40$ and $I=0.77$ (see Sect.~\ref{sec: pri QSP}).

In Sect.~\ref{sec: post trunc} we compared the mean temperature
value obtained for a threshold $I_0=0.6$ and for $I_0=1.0$ in an outer
region (i.e. between 0.4 and 0.5 of $R_{180}$).
In this ring the mean value for the indicator $I$ is 1.14, thus the
expected bias related to the temperature estimator is $\approx$~3\%
(see Sect.~\ref{sec: pri CV}).
We measured a $4\% \pm 3\%$ temperature discrepancy, which is consistent
with the expected bias.
As pointed out in Sect.~\ref{sec: post trunc}, the discrepancy could
also be due to small imperfections in our background model; we are not
able to quantify the amount of this contribution, but we expect it to be
small when considering $I > 0.6$.

To summarize, in external regions our measurements of the cluster
temperature are affected by systematic effects, which depends on the
radius through the factor $I$, i.e. the source-to-background count rate
ratio.
For each ring, we calculate the mean value for $I$, estimate the expected
bias from simulations, and apply a correction to our mean profile.
The expected bias is negligible for internal rings out to 0.30~$R_{180}$
(for which $I \gtrsim 3$), is 2-3\% for 0.30-0.36 and 0.36-0.45 bins, and
is $\approx 5\%$ for the last two bins (i.e. 0.45-0.54 and 0.54-0.70).
We associate to our correction an uncertainty of the same order of the
correction itself, accounting for our limited knowledge from our ``a
posteriori'' tests of the precise value of the bias.
\begin{figure}
  \centering
  \resizebox{80mm}{!}{\includegraphics{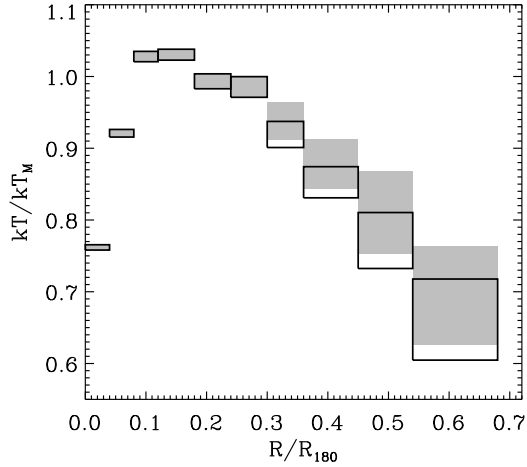}} \\
  \caption{Mean temperature profile rescaled by $R_{180}$ and
  $\mathrm{k}T_\mathrm{M}$. For each ring, empty boxes and shaded regions
  indicate one~sigma uncertainties respectively before and after the bias
  correction.}
  \label{fig: corr prof}
\end{figure}
\begin{table}
  \caption{Mean temperature values rescaled by $\mathrm{k}T_\mathrm{M}$
  and corrected for the biases discussed in the text, for each interval
  in units of $R_{180}$.}
  \label{tab: corr prof}
  \centering
  \begin{tabular}{ll}
    \hline \hline
    Ring$^a$ & Temperature$^b$ \\
    \hline
    0.00-0.04 & 0.762$\pm$0.004 \\
    0.04-0.08 & 0.921$\pm$0.005 \\
    0.08-0.12 & 1.028$\pm$0.007 \\
    0.12-0.18 & 1.030$\pm$0.008 \\
    0.18-0.24 & 0.993$\pm$0.010 \\
    0.24-0.30 & 0.985$\pm$0.014 \\
    0.30-0.36 & 0.938$\pm$0.026 \\
    0.36-0.45 & 0.878$\pm$0.035 \\
    0.45-0.54 & 0.810$\pm$0.058 \\
    0.54-0.70 & 0.694$\pm$0.069 \\
    \hline
  \end{tabular}
  \begin{list}{}{}
    \item[Notes:] $^a$ in units of $R_{180}$; $^b$ in units of
    $\mathrm{k}T_\mathrm{M}$.
  \end{list}
\end{table}
In Fig.~\ref{fig: corr prof} we show the mean temperature profile
before and after the correction for the bias.
In Table~\ref{tab: corr prof} we report for each bin the corrected values;
the uncertainty is the quadrature sum of the statistical error and of the
error associated to our correction.
Hereafter, we will consider the mean profile corrected for the bias,
unless otherwise stated.
Note that the bias is always comparable with the statistical uncertainties.
For this reason, ours can be considered as a definitive work, for what
concerns the measurement of radial temperature profiles of galaxy clusters
with \emph{XMM-Newton}.
We have reached the limits imposed by the instrument and by the analysis
technique, so that further increasing of the number of objects will not
improve the quality of the measurement.


\section{The mean temperature profile} \label{sec: mean prof}

\subsection{Characterizing the profile} \label{sec: best fit}
We fit profiles (see Fig.~\ref{fig: all prof}) beyond 0.2~$R_{180}$
with a linear model and a power law to characterize the profile decline.
By using a linear model
\begin{equation}
\frac{\mathrm{k}T}{\mathrm{k}T_\mathrm{M}} = \mathrm{A} - \mathrm{B}
   \left( \frac{R}{R_{180}} - 0.2 \right)
\end{equation}
we find $\mathrm{A} = 1.02 \pm 0.01$ and $\mathrm{B} = 0.77 \pm 0.11$;
by using a power law
\begin{equation}
\frac{\mathrm{k}T}{\mathrm{k}T_\mathrm{M}} = \mathrm{N}
\left(\frac{R}{0.2 \; R_{180}}\right)^{-\mu}
\end{equation}
we find $\mathrm{N} = 1.03 \pm 0.01$ and $\mu = 0.24 \pm 0.04$.
If the gas can be approximated by a polytrope, we can derive its
index, $\gamma$, from the slope of projected temperature profiles,
$\mu$ \citep{grandi02}:
\begin{equation}
\gamma = 1+\mu/2,
\end{equation}
under the assumption that, at large radii, three-dimensional gas
temperature and density profiles be well described, respectively,
by a power law and a $\beta$-model with $\beta = 2/3$.
For $R > 0.2 \; R_{180}$, we measure $\gamma=1.12\pm0.02$, which is
an intermediate value between those associated to isothermal
($\gamma=1.0$) and adiabatic ($\gamma=1.67$) gas.
\begin{figure}
  \centering
  \resizebox{80mm}{!}{\includegraphics{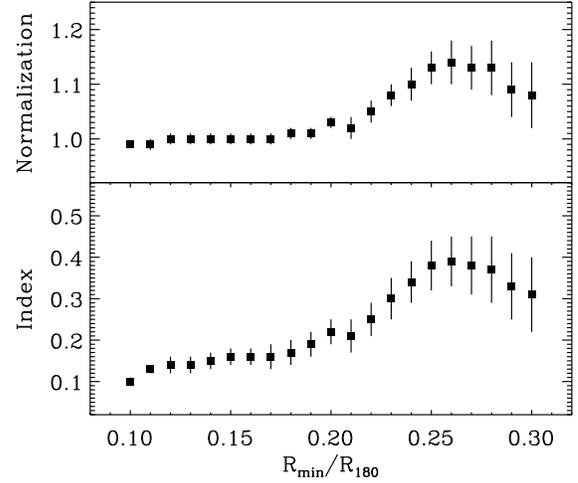}} \\
  \caption{Power-law best-fit parameters obtained by fitting profiles
  beyond a variable radius, $R_\mathrm{min}$, in units of $R_{180}$.
  The normalization is calculated at 0.2. The index best-fit value is
  not constant with $R_\mathrm{min}$, thus the ICM cannot be considered
  as a polytrope.}
  \label{fig: ev par}
\end{figure}
However, we note that the power-law best-fit parameters depend on
the chosen region (see Fig.~\ref{fig: ev par}), as well as the derived
$\gamma$, thus the above values should be taken with some caution.

\subsection{Redshift evolution} \label{sec: evolution}
We divide our clusters in four groups according to the redshift,
to investigate a possible evolution of temperature profiles with
cosmic time.
\begin{figure}
  \centering
  \resizebox{80mm}{!}{\includegraphics{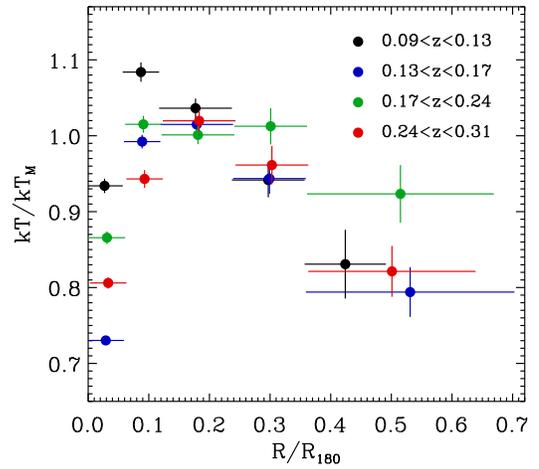}} \\
  \caption{Mean temperature profiles for the four $z$-binned
  groups of clusters. There is no indication of profile evolution.
  The radii have been slightly offset in the plot for clarity.}
  \label{fig: redsh}
\end{figure}
In Fig.~\ref{fig: redsh} we report the mean temperature profiles
for the four groups.
Spectra are fitted in the 0.7-10.0~keV band and $I_0=0.6$ (see
Sect.~\ref{sec: profiles}).
As in the following Sects.~\ref{sec: CCNCC} and
\ref{sec: subsample}, when dividing clusters in subsamples, the
profiles are not corrected for biases (see Sect.~\ref{sec: syst budget}),
because when comparing subsamples we are not interested in
determining the absolute value of the temperature, but in
searching for relative differences.
Moreover, in Fig.~\ref{fig: redsh} and in Fig.~\ref{fig: CC} we
choose larger bin sizes to reduce the error bars (as in
Fig.~\ref{fig: pri QSP}).
The four profiles are very similar: the discrepancy in the outer
regions is comparable to statistical and systematic errors,
the difference in the central region is due to a different
fraction of cool core clusters.
\begin{figure}
  \centering
  \resizebox{80mm}{!}{\includegraphics{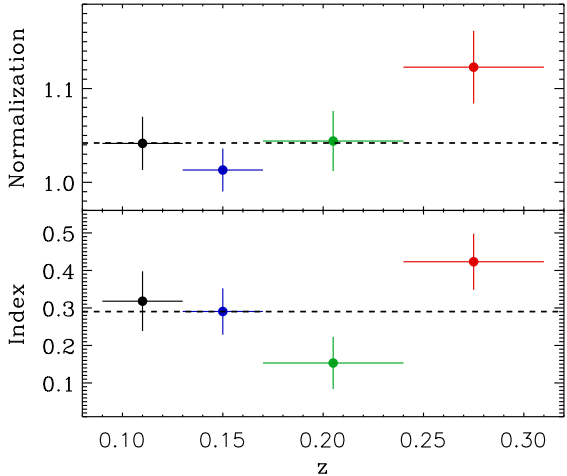}} \\
  \caption{Best fit parameters obtained by fitting
  each group of profiles with a power law beyond 0.2~$R_{180}$.
  The normalization is calculated at 0.2~$R_{180}$. The dashed
  lines indicate the best fit values for the whole sample. No
  clear correlation is found between power-law parameters and
  the redshift.}
  \label{fig: redsh fit}
\end{figure}
We fit each group of profiles with a power law beyond 0.2~$R_{180}$
and report results in Fig.~\ref{fig: redsh fit}.
Since there is no clear correlation between the two parameters
and the redshift, we conclude that from the analysis of our sample
there is no indication of profile evolution up to $z=0.3$.

\subsection{Cool core and non cool core clusters} \label{sec: CCNCC}
In Sect.~\ref{sec: profiles} we defined three groups: clusters
that clearly host a cool core, clusters with no evidence of a
cool core, and uncertain clusters.
\begin{figure}
  \centering
  \resizebox{80mm}{!}{\includegraphics[angle=0]{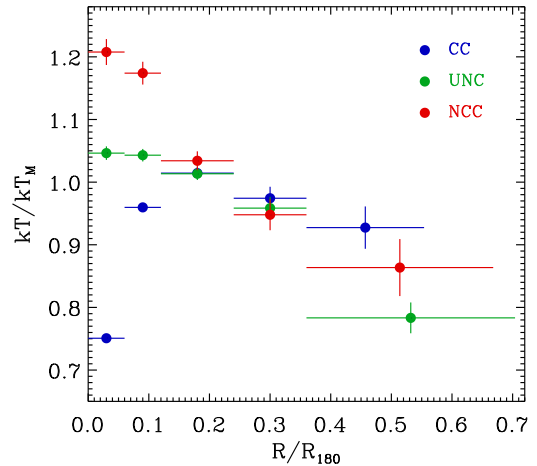}} \\
  \caption{Mean temperature profiles for cool core (blue), non
  cool core (red), and uncertain (green) clusters. Profiles
  differ by definition in the core region and are consistent in
  the outer regions.}
  \label{fig: CC}
\end{figure}
In Fig.~\ref{fig: CC} we show mean temperature profiles for the
three groups.
Spectra are fitted in the 0.7-10.0~keV band and $I_0=0.6$.
Profiles differ by definition in the core region and are
consistent beyond $\approx$~0.1~$R_{180}$.

\subsection{REFL04 and LP07 subsamples} \label{sec: subsample}
Our sample is not complete with respect to any property.
However, most of our clusters ($\approx 2/3$) belong to the
REFLEX Cluster Survey catalog \citep{bohr04}, a statistically
complete X-ray flux-limited sample of 447 galaxy clusters,
and a dozen objects belong to the \emph{XMM-Newton} Legacy
Project sample \citep{pratt07}, which is representative of
an X-ray flux-limited sample with $z<0.2$ and
$\mathrm{k}T>2$~keV.
We then select two subsamples from our sample: clusters
that belong to the REFLEX catalog (REFL04 subsample) and to
the Legacy Project sample (LP07 subsample).
The smaller (i.e. the LP07) is derived from Pratt's parent
sample, by applying our selection criteria based on cluster
temperature and redshift.
We also exclude cluster observations that are heavily affected
by soft proton contamination, however the latter selection
should be equivalent to a random choice and introduce no bias.
Thus, we expect the LP07 subsample to be representative of an
X-ray flux-limited sample of galaxy clusters with $0.1<z<0.2$
and $\mathrm{k}T>3.3$~keV.
The larger (i.e. the REFL04) subsample includes the LP07 one.
Clusters that belong to the REFL04, but not to the LP07, were
observed with \emph{XMM-Newton} for different reasons, they are
not part of a large program and almost all observations have
different PIs.
Thus, there are no obvious reasons to believe that the sample
is significantly biased with respect to any fundamental cluster
property.
A similar reasoning leads to the same conclusion for our whole
sample.

\begin{figure}
  \centering
  \resizebox{80mm}{!}{\includegraphics[angle=0]{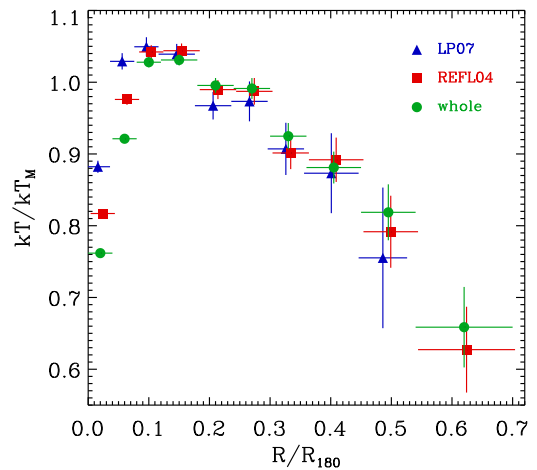}} \\
  \caption{Mean temperature profiles obtained from the LP07
  subsample (blue triangles), the REFL04 subsample (red squares)
  and the whole sample (green circles). The three profiles are
  fully consistent in the outer regions. The radii have been
  slightly offset in the plot for clarity.}
  \label{fig: subsample}
\end{figure}
In Fig.~\ref{fig: subsample} we compare mean temperature profiles
obtained from the two subsamples and the whole sample.
The three profiles are fully consistent beyond $\approx$~0.1~$R_{180}$,
the difference in the central region is due to a different fraction
of CC clusters.
These results allow us to conclude that our whole sample is
representative of hot, intermediate redshift clusters with respect
to temperature profiles, i.e. the quantity we are interested in.

\subsection{Comparison with hydrodynamic simulations} \label{sec: comp sim}
In this subsection we compare our mean temperature profile
with that derived from cluster hydrodynamic simulations by
\cite{borgani04} (hereafter B04).
The authors used the TREE+SPH code GADGET \citep{springel01} to
simulate a concordance cold dark matter cosmological model
($\Omega_\mathrm{m} = 0.3$, $\Omega_\Lambda = 0.7$,
$\sigma_8 = 0.8$, and $h = 0.7$) within a box of 192~$h^{-1}$~Mpc
on a side, 480$^3$ dark matter particles and as many gas
particles.
The simulation includes radiative cooling, star formation and
supernova feedback.
Simulated cluster profiles are scaled by the emission weighted
global temperature and $R_{180}$ calculated from its
definition (i.e. the radius encompassing a spherical density
contrast of 180 with respect to the critical density).
\begin{figure}
  \centering
  \resizebox{80mm}{!}{\includegraphics{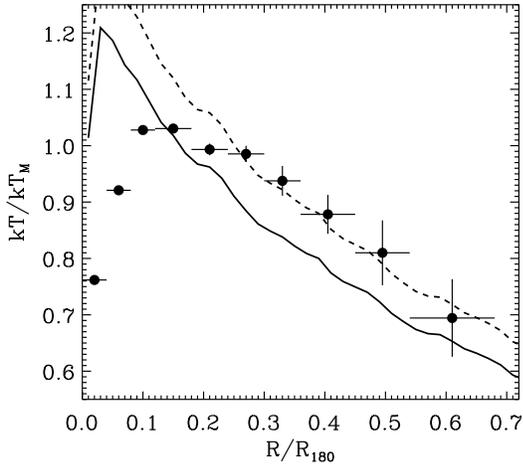}} \\
  \caption{Comparison between our observed mean profile (circles)
  and that derived from hydrodynamic simulations \citep{borgani04}
  by averaging over clusters with $\mathrm{k}T > 3$~keV (solid
  line). The dashed line is obtained by rescaling the solid one
  by 10\%.}
  \label{fig: cfr sim}
\end{figure}
In Fig.~\ref{fig: cfr sim} we compare our
observed profile to the projected mean profile obtained by
averaging over simulated clusters with $\mathrm{k}T > 3$~keV.
The evident mismatch between the two profiles is most likely due
to a different definition for the scaling temperature: actually
it is known that the emission weighted temperature is higher
than the mean temperature obtained from observational data
\citep{mazzotta04}.
By rescaling the B04 profile by 10\%, we find a good agreement
between simulation and our data beyond $\approx$~0.25~$R_{180}$.
Conversely in the core region, simulations are not able to
reproduce the observed profile shape.

\subsection{Comparison with previous observations} \label{sec: comp obs}
In this subsection we compare our mean temperature profile
(LM08) with those obtained by other authors, namely
\cite{grandi02}, \cite{vikh05}, and \cite{pratt07}.
\citeauthor{grandi02} (DM02) have analyzed a sample of 21
hot ($\mathrm{k}T > 3.3$~keV), nearby ($z \lesssim 0.1$) galaxy
clusters observed with \emph{BeppoSAX}.
Their sample includes both CC and NCC clusters.
\citeauthor{vikh05} (V05) have analyzed a sample of 13
nearby ($z \lesssim 0.2$), relaxed galaxy clusters and
groups observed with \emph{Chandra}.
We select from their sample only the hottest ($\mathrm{k}T > 3.3$~keV)
8 clusters, for a more appropriate comparison with our
sample.
\citeauthor{pratt07} (P07) have analyzed a sample of 15 hot
($\mathrm{k}T > 2.8$~keV), nearby ($z \lesssim 0.2$) clusters observed
with \emph{XMM-Newton}.
Clusters of their sample present a variety of X-ray morphology.

Comparing different works is not trivial.
Cluster physical properties, instrumental characteristics,
and data analysis procedures may differ.
Moreover, each author uses his own recipe to calculate a mean
temperature and to derive a scale radius.
We have rescaled temperature profiles obtained by other authors,
by using the standard cosmology (see Sect.~\ref{sec: intro}) and
calculating the mean temperature, $\mathrm{k}T_\mathrm{M}$,
and the scale radius, $R_{180}$, as explained in
Sect.~\ref{sec: profiles}; the aim is to reduce as much as
possible all inhomogeneities.

\begin{figure}
  \centering
  \resizebox{80mm}{!}{\includegraphics{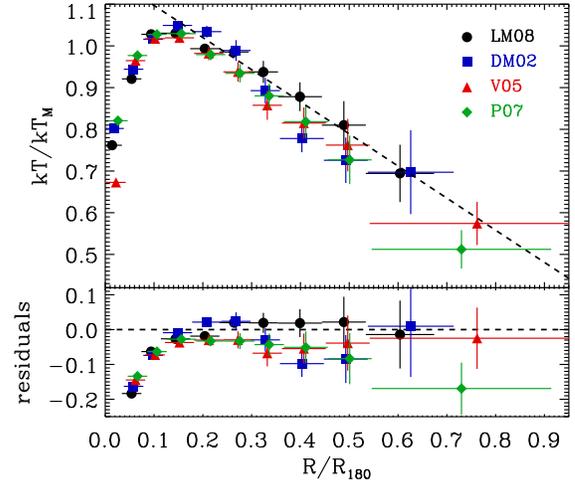}} \\
  \caption{Upper panel: mean temperature profiles obtained from
  this work (black circles, LM08), by \citeauthor{grandi02}
  (blue squares, DM02), by \citeauthor{vikh05} (red upward
  triangles, V05), and by \citeauthor{pratt07} (green
  diamonds, P07). All profiles are rescaled by
  $\mathrm{k}T_\mathrm{M}$ and $R_{180}$ as defined in
  Sect.~\ref{sec: profiles}. The dashed line shows the best
  fit with a linear model beyond 0.2~$R_{180}$ (see
  Sect.~\ref{sec: best fit}) and is drawn to guide the eye.
  Lower panel: residuals with respect to the linear model.
  The LM08 profile is the flattest one.}
  \label{fig: cfr obs}
\end{figure}
\begin{figure}
  \centering
  \resizebox{80mm}{!}{\includegraphics{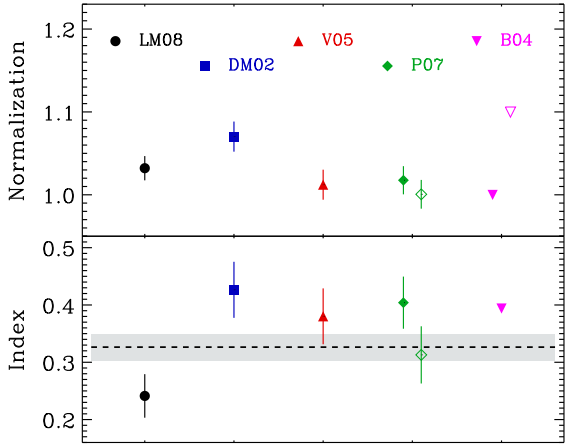}} \\
  \caption{Best fit parameters, obtained by fitting with a
  power law observed and simulated cluster profiles, beyond
  0.2~$R_{180}$: in the upper panel we report the
  normalization, in the lower the index. We use the same
  symbols as in Fig.~\ref{fig: cfr obs} for observed
  clusters and a violet downward triangle for Borgani's
  work (B04). The normalization is calculated at
  0.2~$R_{180}$. For P07 we report two values, empty
  diamonds indicate index and normalization obtained when
  excluding the two outermost measurements (see text for details).
  The empty downward triangle indicates the normalization
  of the B04 rescaled profile (see Sect.~\ref{sec: comp sim}).
  In the lower panel, the dashed line and the shaded region
  represent the weighted average and its one sigma confidence
  interval derived from the observed profiles only (for P07
  we use the lower value, i.e. the empty diamond).
  As previously noted from Fig.~\ref{fig: cfr obs}, the LM08
  profile is the flattest one, but all indices of observed
  profiles are consistent within two sigma.
  Conversely, the B04 profile seems to be significantly
  steeper, but in this case we are not able to provide an
  estimate of parameter uncertainty.}
  \label{fig: cfr pow obs}
\end{figure}

In Fig.~\ref{fig: cfr obs} we compare the four mean temperature
profiles, rescaled by $\mathrm{k}T_\mathrm{M}$ and $R_{180}$.
Due to the correction for the biases described in
Sect.~\ref{sec: syst budget}, our mean profile is somewhat flatter
than others beyond $\approx$~0.2~$R_{180}$.
Discrepancies in the core region are due to a different fraction
of CC clusters.
The outermost point of the P07 profile is $\approx$~25\% lower,
however it is constrained only by two measurements beyond
$\approx$~0.6~$R_{180}$.
Our indicator, $I$, (see Sect.~\ref{sec: profiles}) warns about
the reliability of these two measurements, for which $I \approx$~0.3,
i.e. a half of our threshold, $I_0 = 0.6$.
In Fig.~\ref{fig: crate ratio} we showed that, when using our
analysis technique, lower values of $I$ are associated to a bias
on the temperature measurement.
We assume that a somewhat similar systematic may affect the P07
analysis technique too.
When excluding these two measurements, the P07 mean profile only
extends out to $\approx$~0.6~$R_{180}$ and is consistent with
ours (see also Fig.~\ref{fig: cfr pow obs}).
It is possible that also measurements obtained with other experiments
be affected by a similar kind of systematics, which make the
profiles steeper.

We fit observed and simulated cluster profiles with a power
law beyond 0.2~$R_{180}$ and in Fig.~\ref{fig: cfr pow obs}
report best fit parameters.
The LM08 profile is the flattest one, however all observed
profile indices are consistent within 2-3 sigma.
In Sect.~\ref{sec: syst budget} we have quantified the systematic
underestimate on the temperature measurement associated to our
procedure.
Since it depends on the indicator $I$, which itself depends on the
radius, we expect a net effect also on the profile index, $\mu$,
namely we expect $\mu$ to be overestimated.
For this reason, it is possible that the discrepancy between indices
obtained from different works (reported in Fig.~\ref{fig: cfr pow obs})
may not have a purely statistical origin.
We calculate an average profile index, $\mu = 0.31 \pm 0.02$,
which is significantly lower than that obtained from the B04
profile, $\mu = 0.39$; however, for the simulation we are not able
to provide an estimate of parameter uncertainty.


\section{Summary and conclusions} \label{sec: concl}
We have analyzed a sample of $\approx 50$ hot, intermediate redshift
galaxy clusters (see Sect.~\ref{sec: sample}) to measure their
radial properties.
In this paper we focused on the temperature profiles and postpone
the analysis of the metallicity to a forthcoming paper
(Leccardi \& Molendi 2008, in preparation).
In Sect.~\ref{sec: subsample} we showed that our sample should be
representative of hot, intermediate redshift clusters, at least with
respect to the temperature profile.

Our main results are summarized as follows:
\begin{itemize}
   \item the mean temperature profile declines with radius in the
	 0.2~$R_{180}$-0.6~$R_{180}$ range (see Sect.~\ref{sec: profiles});
   \item when excluding the core region, the profiles are characterized by an
         intrinsic dispersion (6\%) comparable to the estimated systematics,
	 (see Sect.~\ref{sec: profiles});
   \item there is no evidence of profile evolution with redshift out to $z
	 \approx 0.3$ (see Sect.~\ref{sec: evolution});
   \item the profile slope in the outer regions is independent of the presence
         of a cool core (see Sect.~\ref{sec: CCNCC});
   \item the slope of our mean profile is broadly similar to that obtained from
         hydrodynamic simulations, we find a discrepancy of $\approx 10\%$ in
	 normalization probably due to a different definition for the scaling
	 temperature (see Sect.~\ref{sec: comp sim});
   \item when compared to previous works, our profile is somewhat flatter
         (see Sect.~\ref{sec: comp sim}), probably due to a different level
         of characterization of systematic effects, which become very
	 important in the outer regions.
\end{itemize}

The above results have been obtained using a novel data analysis technique,
which includes two major improvements.
Firstly, we used the background modeling, rather than the background
subtraction, and the Cash statistic rather than the $\chi^2$; this method
requires a careful characterization of all background components.
Secondly, we assessed in details systematic effects.
We performed two groups of test: prior to the analysis, we made use of
extensive simulations to quantify the impact of different components on
simulated spectra; after the analysis, we investigated how the measured
temperature profile changes, when choosing different key parameters.

From a more general point of view, ours is an attempt to measure cluster
properties, as far out as possible, with EPIC instruments.
Perhaps, the most important justification for our efforts in this direction
is that, for the next 5-10 years, there will be no experiments with
comparable or improved capabilities, as far as low surface brightness
emission is concerned.


\begin{acknowledgements}
We acknowledge the financial contribution from contract ASI-INAF I/088/06/0,
I/023/05/0, and I/088/06/0.
We thank S.~Ghizzardi, M.~Rossetti, and S.~De~Grandi for a careful reading
of the manuscript.
We thank S.~Borgani, G.~W.~Pratt, and A.~Vikhlinin for kindly providing their
temperature profiles.
\end{acknowledgements}

\bibliographystyle{aa}
\bibliography{paper2}

\appendix

\section{The analysis of ``closed'' observations} \label{sec: closed}
We have analyzed a large number ($\approx$~50) of observations
with the filter wheel in the ``closed'' position to characterize in
detail the EPIC-MOS internal background and to provide constraints to
the background model, which we use for analyzing our data.
Exposure times of individual observations span between 5 and 100~ks
for a total exposure time of $\approx$~650~ks.

\begin{figure}
  \centering
  \resizebox{80mm}{!}{\includegraphics[angle=270]{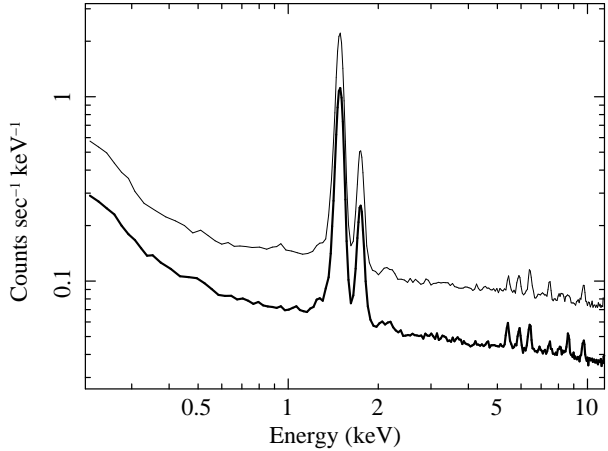}} \\
  \caption{MOS1 (thick) and MOS2 (thin) spectra from closed
  observations in the whole energy band, i.e. 0.2-11.3~keV. MOS2
  spectrum is scaled by a factor of 2 for clarity. Spectra are
  accumulated in the 10$^\prime$-12$^\prime$ ring. The total
  exposure time is $\approx$~650~ks.}
  \label{fig: NXB1}
\end{figure}

For each observation, we select 6 concentric rings
(0$^\prime$-2.75$^\prime$, 2.75$^\prime$-4.5$^\prime$,
4.5$^\prime$-6$^\prime$, 6$^\prime$-8$^\prime$,
8$^\prime$-10$^\prime$, and 10$^\prime$-12$^\prime$) centered on the
detector center.
For each instrument (i.e. MOS1 and MOS2) and each ring, we produce
the total spectrum by summing, channel by channel, spectral counts
accumulated during all observations.
The appropriate RMF is associated to each total spectrum and a
minimal grouping is performed to avoid channels with no counts.
In Fig.~\ref{fig: NXB1} we report the total spectra accumulated in
the 10$^\prime$-12$^\prime$ ring, for MOS1 and MOS2, in the
0.2-11.3~keV band.
Closed observation events are solely due to the internal background,
which is characterized by a cosmic-ray induced continuum (NXB) plus
several fluorescence emission lines.
The most intense lines are due to Al ($\approx$~1.5~keV) and Si
($\approx$~1.8~keV).
Beyond 2~keV we fit the NXB with a single power law (index 0.24 and
0.23 for MOS1 and MOS2 respectively); instead, for the 0.7-10.0~keV
range, a broken power-law (see Table~\ref{tab: NXB}) is more
appropriate.
Emission lines are modeled by Gaussians.
Note that particle background components are not multiplied by the
effective area.
\begin{table}
  \caption{Best fit parameters for the NXB broken power law.
  $\Gamma_1$ and $\Gamma_2$ are the slopes below and above the break
  energy, $E_\mathrm{B}$.}
  \label{tab: NXB}
  \centering
  \begin{tabular}{cccc}
    \hline \hline
     & $\Gamma_1$ & $E_\mathrm{B}$ [keV] & $\Gamma_2$ \\
    \hline
    MOS1 & 0.22 & 7.0 & 0.05 \\
    MOS2 & 0.32 & 3.0 & 0.22 \\
    \hline
  \end{tabular}
\end{table}

In Table~\ref{tab: lines} we list the emission lines of our model
with their rest frame energies.
\begin{table}
  \caption{Instrumental emission lines in the 0.7-10.0~keV energy band.}
  \label{tab: lines}
  \centering
  \begin{tabular}{lclc}
    \hline \hline
    Line & E [keV] & Line & E [keV] \\
    \hline
    Al~K$\alpha$   & 1.487 & Mn~K$\beta$    & 6.490 \\
    Al~K$\beta$    & 1.557 & Fe~K$\beta$    & 7.058 \\
    Si~K$\alpha$   & 1.740 & Ni~K$\alpha$   & 7.472 \\
    Si~K$\beta$    & 1.836 & Cu~K$\alpha$   & 8.041 \\
    Au~M$\alpha$   & 2.110 & Ni~K$\beta$    & 8.265 \\
    Au~M$\beta$    & 2.200 & Zn~K$\alpha$   & 8.631 \\
    Cr~K$\alpha$   & 5.412 & Cu~K$\beta$    & 8.905 \\
    Mn~K$\alpha$   & 5.895 & Zn~K$\beta$    & 9.572 \\
    Cr~K$\beta$    & 5.947 & Au~L$\alpha$   & 9.685 \\
    Fe~K$\alpha$   & 6.400 & & \\
    \hline
  \end{tabular}
\end{table}
Normalization values are always reported in XSPEC units.
Lines are determined by 3 parameters: peak energy, intrinsic width
and normalization.
The energy of Al~K$\alpha$, E$_\mathrm{Al}$, is free to allow for a
small shift in the energy scale; the energies of Al, Si, and Au-M
lines are linked to E$_\mathrm{Al}$ in such a way that a common shift,
$\Delta$E, is applied to all lines.
Similarly, the energy of Cr~K$\alpha$, E$_\mathrm{Cr}$, is free and
the energies of all other lines are linked to E$_\mathrm{Cr}$.
The intrinsic width is always fixed to zero, except for Al and Si
lines for which it is fixed to 0.0022~keV to allow for minor
mismatches in energy calibrations for different observations.
Normalizations of K$\alpha$, Al, and Si lines are free, while
normalizations of K$\beta$ lines are forced to be one seventh of
the correspondent K$\alpha$ line \citep{keith78}.
The correlation between broken power-law and Gaussian parameters is
very weak.

As noticed by \citet{kuntz06}, there are observations in which the
count rate of some CCDs is very different, especially at low energies,
indicating that the NXB spectral shape is not constant over the detector.
In particular, this problem affects MOS1 CCD-4 and CCD-5 and MOS2 CCD-2
and CCD-5.
Since our procedure requires background parameters to be rescaled from
the outer to the inner rings, we always exclude the above mentioned
``bright'' CCDs from data analysis, when using the 0.7-10.0~keV band
(see Sect.~\ref{sec: prelim}).
This is not necessary when using the band above 2~keV, because the
effect is negligible for almost all observations.

\begin{figure*}
  \centering
  \resizebox{180mm}{!}{\includegraphics[angle=0]{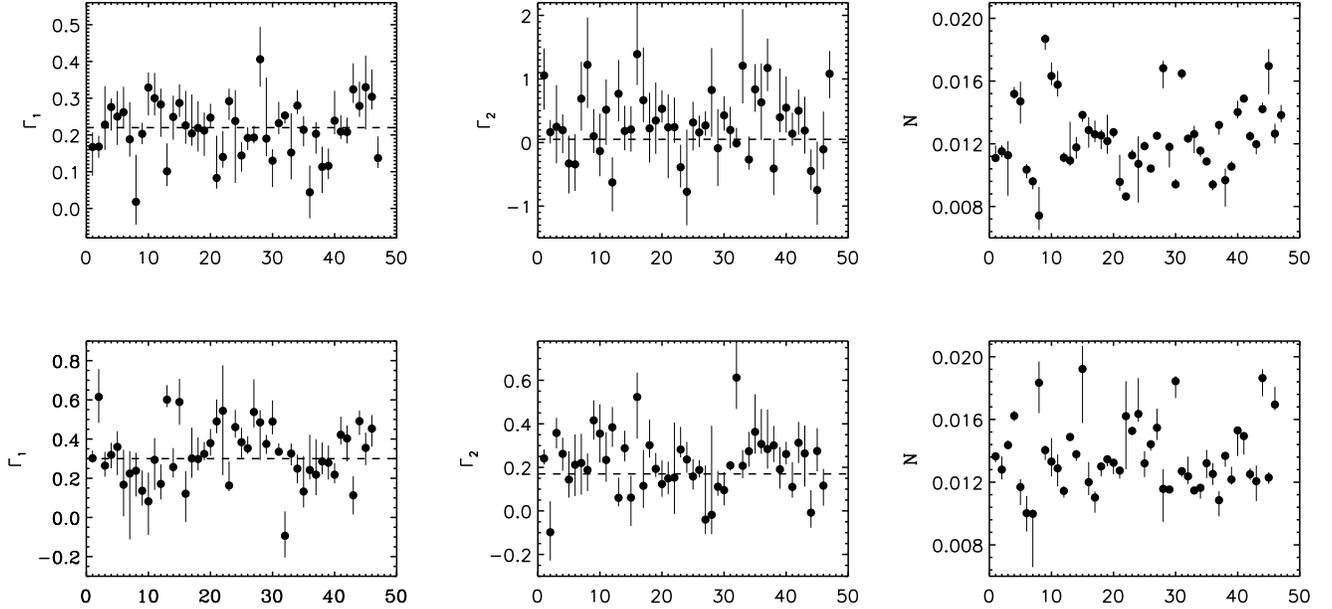}} \\
  \caption{$\Gamma_1$, $\Gamma_2$, and $N$ values for MOS1 (top) and
  MOS2 (bottom) for all closed observations analyzed. The dotted lines
  are the best fit values reported in Table~\ref{tab: NXB}. For
  $\Gamma_1$ and $\Gamma_2$ the scatter is comparable with the
  uncertainties, while for $N$ there is an intrinsic scatter of
  $\approx$~20\%. $N$ values are reported in XSPEC units.}
  \label{fig: tvar}
\end{figure*}

After the exclusion of the bright CCDs, we fit spectra accumulated in
the 10$^\prime$-12$^\prime$ ring for different closed observations, to
check for temporal variations of the NXB.
In Fig.~\ref{fig: tvar} we report the values of broken power-law free
parameters (namely the slopes, $\Gamma_1$ and $\Gamma_2$, and the
normalization, $N$) for both MOS in the 0.7-10.0~keV band.
The scatter of $\Gamma_1$ and $\Gamma_2$ values is of the same order
of magnitude as the statistical uncertainties, while the scatter of
the $N$ values ($\approx$~20\%) is not purely statistic, i.e. NXB
normalization varies for different observations.

\begin{figure}
  \centering
  \resizebox{80mm}{!}{\includegraphics[angle=0]{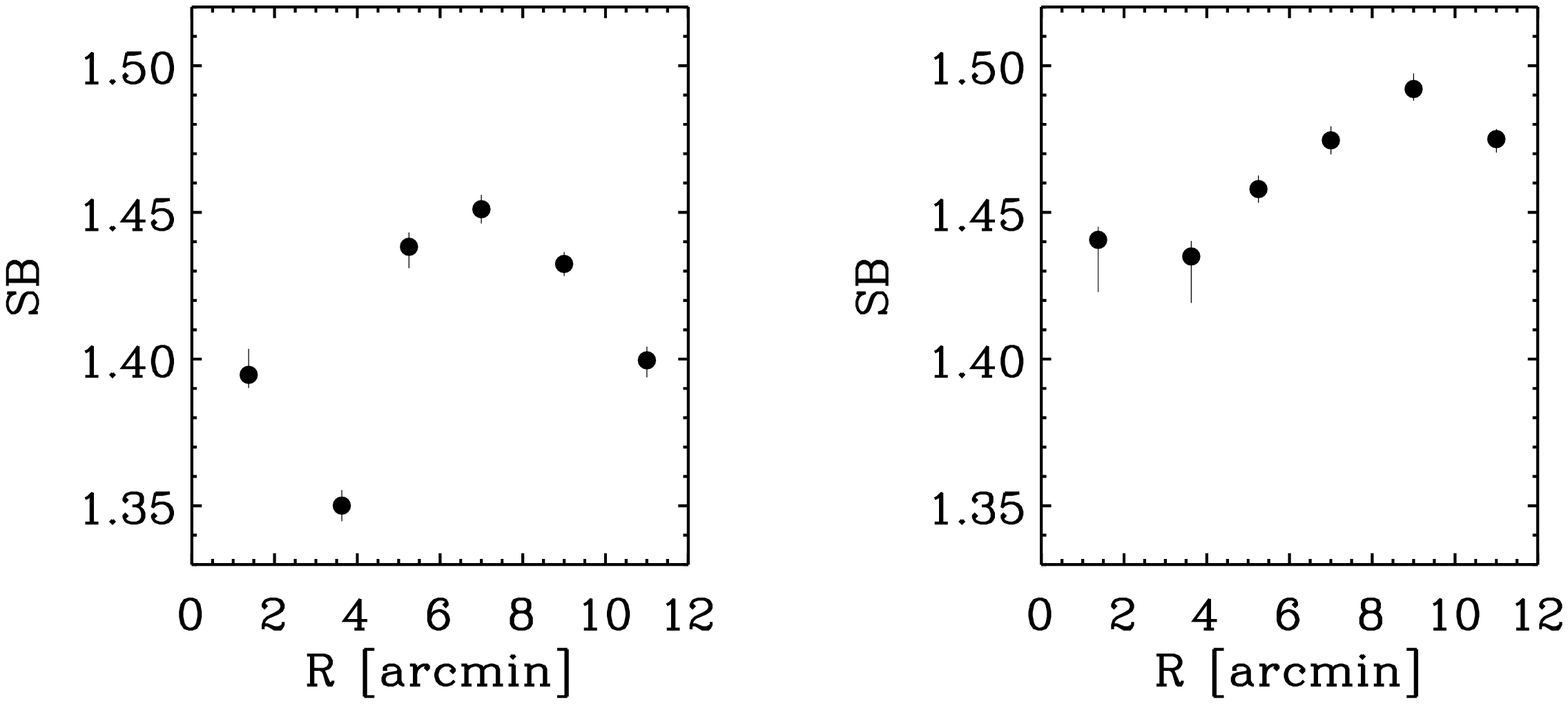}} \\
  \caption{Surface brightness best fit values for MOS1 (left) and MOS2
  (right) as a function of the distance from the detector center.}
  \label{fig: spvar}
\end{figure}

We also check for spatial variations of the internal background.
As explained at the beginning of this section, we accumulate the total
spectrum for each of the 6 rings and for each instrument.
We define the surface brightness, \textit{SB}, as the ratio between $N$
and the area of the ring.
In Fig.~\ref{fig: spvar} we report MOS1 and MOS2 best fit values of
\textit{SB} as a function of the distance from the center, by fixing
$\Gamma_1$ and $\Gamma_2$.
The spatial variations are greater than statistical errors but smaller
than 5\%.
To a first approximation, the NXB is flat over the detector.
We find similar results, both in terms of temporal and spatial
variations, when fitting spectra above 2~keV.

Emission lines show rather weak temporal variations and most of them
(namely all except for Al, Si, and Au) have a uniform distribution
over the detector.
Al lines are more intense in the external CCDs, while Si lines are
more intense in the central CCD.
Conversely, Au lines are very localized in the outer regions of the
field of view, thus we model them only when analyzing rings beyond
3.5$^\prime$.

\section{The analysis of ``blank field'' observations} \label{sec: blankfield}
A large number ($\approx$~30) of ``blank field'' observations have been
analyzed to characterize the spectrum of other background components.
Exposure times of individual observations span between 30 and 90~ks for
a total exposure time of $\approx$~600~ks.
Almost each observation has a different pointing in order to maximize
the observed sky region and minimize the cosmic variance of the X-ray
background.

Data are prepared and cleaned as described in Sects.~\ref{sec: prelim}
and \ref{sec: INOUT}.
For each instrument (i.e. MOS1 and MOS2) and each filter (i.e. THIN1 and
MEDIUM), we produce total spectra by summing, channel by channel, spectral
counts accumulated during all observations, after the selection of the
same rings used for closed observations (see Appendix~\ref{sec: closed}).
The appropriate RMF and ARF are associated to each spectrum and a minimal
grouping is performed to avoid channels with no counts.
We also calculate the average $R_\mathrm{SB}$ (see Sect.~\ref{sec: INOUT}),
obtaining 1.09$\pm$0.01 for both filters and both detectors.

\begin{figure}
  \centering
  \resizebox{80mm}{!}{\includegraphics[angle=270]{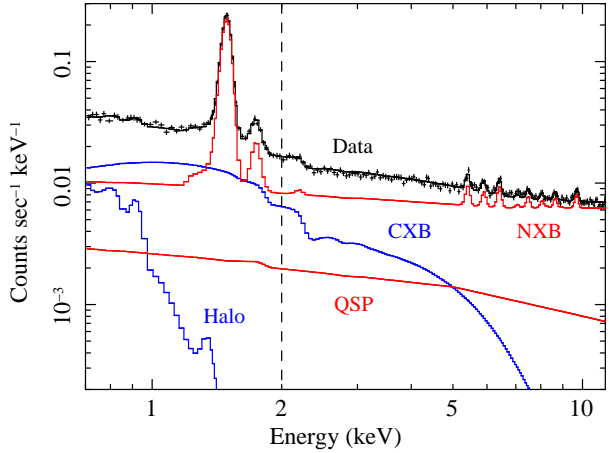}} \\
  \caption{MOS1 spectrum from blank field observations in the
  10$^\prime$-12$^\prime$ ring. Above 2~keV the spectrum is simpler.
}
  \label{fig: bf}
\end{figure}

Inside the field of view, the spectral components are the following (see
Fig.~\ref{fig: bf}):
\begin{itemize}
   \item the X-ray background from Galaxy Halo (HALO),
   \item the cosmic X-ray background (CXB),
   \item the quiescent soft protons (QSP),
   \item the cosmic ray induced continuum (NXB),
   \item the fluorescence emission lines.
\end{itemize}
Only the photon components (i.e. HALO and CXB) are multiplied by the
effective area and absorbed by our Galaxy.
The equivalent hydrogen column density along the line of sight,
$N_\mathrm{H}$, is fixed to the 21 cm measurement \citep{dickey90},
averaged over all fields.
We selected blank field observations pointed at high galactic latitude,
therefore $N_\mathrm{H}$ is $< 10^{21}$~cm$^{-2}$ and the absorption
effect is negligible above 1~keV.

In the 0.7-10.0~keV band, the total model is composed of a thermal component
(HALO), a power law (CXB), two broken power laws (QSP and NXB), and several
Gaussians (fluorescence emission lines).
The thermal model (APEC in XSPEC) parameters are: $\mathrm{k}T=0.197$~keV,
$Z=1.0\;Z_\odot$, and $z=0.0$ \citep{kuntz00}.
The slope of the CXB power law is fixed to 1.4 \citep{deluca04} and the
normalization is calculated at 3~keV to minimize the correlation with the
slope.
The QSP broken power law has a break energy at 5.0~keV; the slopes are fixed
to 0.4 (below 5~keV) and 0.8 (above 5~keV).
The model parameters for the internal background are the same as reported in
Appendix~\ref{sec: closed}.
In the 2.0-10.0~keV band the model is simpler, namely three power laws and
several Gaussians, and more stable.
The HALO component is negligible above 2~keV, the CXB model is the same as
in the 0.7-10.0~keV band, the slope of the QSP power law is fixed to 1.0,
and the model parameters for the internal background are those reported in
Appendix~\ref{sec: closed}.

Most components have rather similar spectral shapes (see Fig.~\ref{fig: bf}),
therefore a high degree of parameter degeneracy is present.
In such cases, it is useful to constrain as many parameters as possible.
Events outside the field of view are exclusively due to the internal
background, therefore the spectrum accumulated in this region provides a
good estimate of the NXB normalization, $N_\mathrm{NXB}$.
By analyzing closed (CL) observations we found that the ratio between
$N_\mathrm{NXB}$ calculated in any two detector regions is independent of
the particular observation:
\begin{equation}
\frac {N_\mathrm{NXB}(R_1;O_1)} {N_\mathrm{NXB}(R_2;O_1)} =
\frac {N_\mathrm{NXB}(R_1;O_2)} {N_\mathrm{NXB}(R_2;O_2)} \, ,
\label{eq: rel NXB}
\end{equation}
where $R_{1,2}$ are two detector regions and $O_{1,2}$ are two observations.
By using the region outside the field of view (OUT), from Eq.~\ref{eq: rel NXB}
we estimate and fix $N_\mathrm{NXB}$ for each ring (R) of blank field (BF)
observations:
\begin{equation}
N_\mathrm{NXB}(R;BF) = N_\mathrm{NXB}(R;CL) \times
\frac {N_\mathrm{NXB}(OUT;BF)} {N_\mathrm{NXB}(OUT;CL)} \, .
\label{eq: rel2 NXB}
\end{equation}

\begin{table}
  \caption{Best fit results for the analysis of blank field observations in the
  10$^\prime$-12$^\prime$ ring.}
  \label{tab: bf results}
  \centering
  \begin{tabular}{ccccc}
    \hline \hline
    Instr. & Filter & $N_\mathrm{HALO}$ & $N_\mathrm{QSP}$ & $N_\mathrm{CXB}^\mathrm{a}$ \\
     & & [$10^{-4}$] & [$10^{-3}$] & [$10^{-2}$] \\
    \hline
    MOS1 & THIN1  & 1.7$\pm$0.1 & 2.4$\pm$0.1 & 5.1$\pm$0.1 \\
    MOS2 & THIN1  & 1.6$\pm$0.1 & 2.5$\pm$0.1 & 5.0$\pm$0.1 \\
    \hline
    MOS1 & MEDIUM & 1.4$\pm$0.1 & 2.6$\pm$0.1 & 6.0$\pm$0.1 \\
    MOS2 & MEDIUM & 1.6$\pm$0.1 & 2.4$\pm$0.1 & 5.8$\pm$0.1 \\
    \hline
  \end{tabular}
  \begin{list}{}{}
    \item[Notes:] $\mathrm{^a}$ calculated at 3~keV.
    \end{list}
\end{table}
In Table~\ref{tab: bf results} we report the best fit values for the
normalization of the HALO, $N_\mathrm{HALO}$, of the QSP, $N_\mathrm{QSP}$,
and of the CXB, $N_\mathrm{CXB}$, in the 10$^\prime$-12$^\prime$ ring, for
MOS1 and MOS2 instruments and for THIN1 and MEDIUM filters.
Spectra are fitted in the 0.7-10.0~keV energy band.
We stress the remarkably good agreement between MOS1 and MOS2, for all
parameters.
Moreover, we point out that, when comparing observations with different
filters, values for $N_\mathrm{HALO}$ and $N_\mathrm{QSP}$ also agree,
while values for $N_\mathrm{CXB}$ are significantly different
($\approx$~20\%) because of the cosmic variance ($\approx$~15\% expected
for the considered solid angles).

By construction (see Eq.~\ref{eq: RSB}) $R_\mathrm{SB}$ is related
to $N_\mathrm{QSP}$ so that the higher $R_\mathrm{SB}$, the higher
$N_\mathrm{QSP}$.
For observations that are not contaminated by QSP, we will measure
$R_\mathrm{SB} \approx 1.0$ and $N_\mathrm{QSP} \approx 0.0$.
Since $R_\mathrm{SB}$ values span a relatively small range (roughly
between 1.0 and 1.5) we approximate the relation between $R_\mathrm{SB}$
and $N_\mathrm{QSP}$ with a linear function:
$N_\mathrm{QSP}=A\times(R_\mathrm{SB}-1)$.
The scaling factor, $A \approx 0.03$, is determined from the analysis of
blank fields observations, for which we have measured
$R_\mathrm{SB}=1.09\pm0.01$ and $N_\mathrm{QSP}=(2.5\pm0.1)\times 10^{-3}$.
Thus, for each observation we model the bulk of the QSP component by
deriving $N_\mathrm{QSP}$ from $R_\mathrm{SB}$ (see
Sects.~\ref{sec: ext ring} and \ref{sec: int rings}).
In Sects.~\ref{sec: pri QSP} and \ref{sec: post INOUT} we discuss possible
systematics related to QSP and show that the linear approximation used
above is satisfactory.

As mentioned in Sect.~\ref{sec: ext ring}, we estimate the normalizations
of the background components in the 10$^\prime$-12$^\prime$ ring and
rescale them in the inner rings; when considering the 0.7-10.0~keV energy
band, a simple rescaling by the area ratio is too rough and causes
systematic errors, especially in the outer regions where cluster emission
and background fluctuations become comparable.
To overcome this problem, we proceed in the following manner:
we fit blank field spectra, by fixing $N_\mathrm{NXB}$ and
$N_\mathrm{QSP}$, and determine $N_\mathrm{CXB}$ and $N_\mathrm{HALO}$
best fit values.
For each ring and instrument, we define a correction factor, $K(r)$:
\begin{equation}
K(r) = \frac{N_\mathrm{obs}}{N_\mathrm{exp}} ,
\end{equation}
where $N_\mathrm{obs}$ is the best fit value that we have just obtained
and $N_\mathrm{exp}$ is derived by rescaling the value measured in the
10$^\prime$-12$^\prime$ ring by the area ratio.
In Table~\ref{tab: corr factor} we report the values for $K(r)$ for
all cases.
\begin{table}
  \caption{Correction factors.}
  \label{tab: corr factor}
  \centering
  \begin{tabular}{lcccc}
    \hline \hline
     Ring & \multicolumn{2}{c}{HALO} & \multicolumn{2}{c}{CXB} \\
     & MOS1 & MOS2 & MOS1 & MOS2 \\
    \hline
    0$^\prime$-2.75$^\prime$   & 0.62 & 0.68 & 0.80 & 0.91 \\
    2.75$^\prime$-4.5$^\prime$ & 0.74 & 0.70 & 0.70 & 0.78 \\
    4.5$^\prime$-6$^\prime$    & 0.63 & 0.65 & 0.89 & 0.95 \\
    6$^\prime$-8$^\prime$      & 0.74 & 0.71 & 0.89 & 0.92 \\
    \hline
  \end{tabular}
\end{table}
$K(r)$ is a second order correction, because the contribution of CXB and
HALO components to the total flux is relatively small: when considering only
the 0.7-2.0~keV band (i.e. the energy range in which these components are
more intense), the HALO-to-total and the CXB-to-total flux ratios are
$\approx$~5\% and $\approx$~20\% respectively.
Thus, the effective correction is only of a few percent for both cases.
Different observations have different centers in detector coordinates
and the intensity of the various components depends on the particular
observation; these facts could cause some discrepancies, however since we
have analyzed a large number of blank and cluster fields, we expect only
a few percent systematic effect on the mean profile.
When considering the band above 2~keV, the statistical quality of the data
is poorer, therefore the rescaling by the area ratio (i.e. no correction
factor) can be considered a good approximation for both CXB and NXB.
The QSP value is rescaled by the soft proton vignetting profile
\citep{kuntz06} and does not require any correction factor.

Unfortunately, a precise characterization of the QSP component for
EPIC-pn is not possible.
Uncertainties on $N_\mathrm{NXB}$ are very large, because the region
outside the EPIC-pn field of view is much smaller than the MOS one; the
presence of a non negligible fraction of out-of-time events introduces
a further complication.
Moreover, the EPIC-pn background is much less stable than the EPIC-MOS one,
especially below 2~keV.
The EPIC-pn instrument has further drawbacks due to the electronic board
near the detector: the NXB spatial distribution is not flat and the
emission due to Ni-Cu-Zn lines (between $\approx$~7.5~keV and
$\approx$~9.5~keV) is more intense in the outer rings.
For these reasons, as mentioned in Sect.~\ref{sec: analysis}, we consider
only EPIC-MOS data in our analysis.

\end{document}